\documentclass[11pt]{article}
\usepackage{amsmath}
\usepackage{graphicx}
\usepackage{amsfonts}
\usepackage{amssymb}
\usepackage{epsfig}
\usepackage{color}
\usepackage{psfrag}
\usepackage{epstopdf}
\usepackage{mathrsfs}
\usepackage{stmaryrd} 

\newcommand{\EQ}{\begin{equation}}
\newcommand{\EN}{\end{equation}}
\newcommand{\be}{\begin{equation}}
\newcommand{\ee}{\end{equation}}
\newcommand{\bea}{\begin{eqnarray}}
\newcommand{\eea}{\end{eqnarray}}

\newcommand{\rd}{{\rm d}}

\usepackage{bm}

\newcommand{\newfig}[1]{\parbox{1.8cm}{\epsfig{file=#1,height=3.4cm}}}

\usepackage{caption}

\begin{document} 
\topmargin 0pt
\oddsidemargin 5mm
\renewcommand{\thefootnote}{\arabic{footnote}}
\newpage
\topmargin 0pt
\oddsidemargin 5mm
\renewcommand{\thefootnote}{\arabic{footnote}}

\newpage

\begin{titlepage}
\begin{flushright}
\end{flushright}
\begin{center}
{\large {\bf Long range correlations generated by phase separation.\\ Exact results from field theory}}\\
\vspace{1.8cm}
{\large Gesualdo Delfino$^{1,\natural}$ and Alessio Squarcini$^{2,3,\flat}$}\\ 
\vspace{0.5cm}
${}^1${\em SISSA -- Via Bonomea 265, 34136 Trieste, Italy}\\
{\em INFN -- sezione di Trieste, Italy}\\
${}^2${\em Max-Planck-Institut f$\ddot{u}$r Intelligente Systeme,\\
Heisenbergstr. 3, D-70569 Stuttgart, Germany}\\
${}^3${\em IV. Institut f$\ddot{u}$r Theoretische Physik, Universit$\ddot{a}$t Stuttgart,\\
Pfaffenwaldring 57, D-70569 Stuttgart, Germany}\\

\end{center}
\vspace{1.2cm}

\renewcommand{\thefootnote}{\arabic{footnote}}
\setcounter{footnote}{0}

\begin{abstract}
\noindent
We consider near-critical planar systems with boundary conditions inducing phase separation. While order parameter correlations decay exponentially in pure phases, we show by direct field theoretical derivation how phase separation generates long range correlations in the direction parallel to the interface, and determine their exact analytic form. The latter leads to specific contributions to the structure factor of the interface.
\end{abstract}

\vspace{.4cm}

\vfill
$^\natural$delfino@sissa.it, $^\flat$squarcio@is.mpg.de

\end{titlepage}

\newpage

\tableofcontents

\section{Introduction}
\label{*sec.01}
The notion of interface is relevant to different areas of physics. In particle physics the simplest model of the confining potential between a quark and an antiquark is obtained seeing them as the endpoints of a string whose time propagation generates a two-dimensional surface; for large separations $r$ the potential then grows as $\sigma r$ due to the surface tension $\sigma$. In a statistical system at phase coexistence, in any dimension $d\geq 2$, suitable boundary conditions induce a separation between different phases which is commonly described in terms of an interface. The connection between the two physical problems becomes explicit when duality relates a lattice gauge theory to a spin model (see e.g. \cite{ID}). 

It is clear that the notion of surface/interface provides an {\it effective} description of phenomena for which one is unable to perform a first principle derivation from the underlying field theory (gauge theory for confinement, field theory of the scaling region for a near-critical statistical system). The process of endowing the interface with fluctuations able to reproduce the observed properties leads to effective string actions \cite{Goto,Nambu,LW} for confinement, and to capillary wave theory \cite{BLS} and its extensions in statistical physics. These approaches account for the presence of long wavelength modes for the effective degrees of freedom, i.e. the deviations of the interface from its average position. In turn, the existence of these long wavelength modes should imply long range correlations in the underlying field theory. Since, referring from now on to phase separation (e.g. in the Ising model slightly below the critical temperature\footnote{We always refer to systems with short range interactions and in their scaling limit, i.e. close to a second order phase transition point. As a consequence, our results are characteristic of a given universality class.}), all correlations decay exponentially in a pure phase, long range correlations of the order parameter must be {\it generated} by phase separation. Although this implication has been pointed out and investigated since long time in the context of inhomogeneous fluids \cite{Wertheim,Weeks,Evans}, these correlations have never been derived within the underlying field theory. In this paper we perform this derivation {\it exactly} in the two-dimensional case. 

We are able to do this because it has been shown in the last few years \cite{DV,DS2,fpu} how phase separation in near-critical two-dimensional systems can be described in a fundamental, general and exact way supplementing with the required boundary conditions the bulk field theory, i.e. the field theory corresponding to the scaling limit of the pure phases. This has allowed, in the first place, to determine the order parameter profiles\footnote{In the Ising case one recovers the exact lattice result of \cite{Abraham_strip,Abraham}.} (one-point functions) and to {\it derive} from them the properties of the interfacial region, including the deviations from the simple curve picture \cite{DV,DS2}. The theory has also been extended to interfaces at boundaries \cite{DS} and to interface localization \cite{localization}.

Here we move on to the determination of the two-point function of the order parameter, in the large distance regime relevant for the issue discussed above. We consider the system in the infinitely long strip $|y|\leq R/2$ in the $xy$ plane, with boundary conditions on the two edges favoring a phase $a$ for $x<0$ and a phase $b$ for $x>0$ (Fig.\ref{fig.01}), with $a$ and $b$ coexisting phases. In the relevant regime $R$ much larger than the bulk correlation length $\xi$, phase separation is exhibited by the variation of the expectation value $\langle\sigma(x,y)\rangle_{ab}$ of the order parameter field $\sigma(x,y)$ (spin field within the magnetic terminology) from $\langle\sigma\rangle_a$ to $\langle\sigma\rangle_b$ as $x$ varies from $-\infty$ to $+\infty$. We denote by $\langle\cdots\rangle_{ab}$ the expectation values for boundary conditions changing from $a$ to $b$ at $x=0$, and by $\langle\cdots\rangle_a$ the expectation values in the pure phase $a$. The analytic formulae of this paper hold for systems for which phase separation takes place in its simplest form, i.e. without the formation of an intermediate macroscopic layer of a third phase; these include the universality classes of the Ising model (which has only two phases), those of the three- and four-state Potts model, and others (see \cite{localization} for a classification). For the two-point function we obtain, in particular,
\bea
 \langle\sigma(x,y)\sigma(x,-y)\rangle_{ab} &=& \left(\frac{\langle\sigma\rangle_a+\langle\sigma\rangle_b}{2}\right)^2+\frac{\langle\sigma\rangle_b^2-\langle\sigma\rangle_a^2}{2}\,\textrm{erf}\left(\sqrt{\frac{2m}{R}}\,x\right)\nonumber\\
&+& \left(\frac{\langle\sigma\rangle_a-\langle\sigma\rangle_b}{2}\right)^2\left(1-\frac{4}{\pi}\sqrt{\frac{2y}{R}}\,\textrm{e}^{-\frac{2m}{R}x^2}\right)+O((y/R)^{3/2})\,
\label{leading}
\eea
in the limits
\EQ
\xi\ll y\ll R/2\,;
\label{limits}
\EN
$m\propto 1/\xi$ coincides with the interfacial tension and $\textrm{erf}(x)=(2/\sqrt{\pi})\int_0^x \textrm{d}u\,\textrm{e}^{-u^2}$ is the error function. It follows from (\ref{leading}) that 
\EQ
\lim_{R\to\infty}\langle\sigma(x,y)\sigma(x,-y)\rangle_{ab}=\frac{\langle\sigma\rangle_a^2+\langle\sigma\rangle_b^2}{2}\,,\hspace{.5cm}y\gg\xi\,,
\label{plane}
\EN
a result explained by the fact that, as recalled in the next section, the horizontal fluctuations of the interface grow as $\sqrt{R}$; hence for $R=\infty$, no matter the value of $x$, one obtains the average of the correlator over the two pure phases; on the other hand, $\langle\sigma(x_1,y_1)\sigma(x_2,y_2)\rangle_a$ tends to $\langle\sigma\rangle_a^2$ for separations much larger than $\xi$. For the Ising model this averaging property is known rigorously for $n$-point functions (see \cite{Abraham}). 

The term proportional to $\sqrt{y/R}$ in (\ref{leading}) is particularly interesting, since it shows that phase separation generates long range (i.e. not exponentially suppressed) correlations in the vertical direction (parallel to the interface). It also shows that, within the limits (\ref{limits}), these correlations {\it grow} as $\sqrt{y}$ for $R$ fixed, and that they vanish for $R=\infty$. 

On the side of effective theories, the characterization of order parameter correlations in presence of phase separation is especially pursued in momentum space (see \cite{BKV,PRWE,HD} and references therein), focusing on the {\it interface structure factor}
\begin{equation}
\label{sf}
{S}(q) = \frac{1}{2({\langle\sigma\rangle_a-\langle\sigma\rangle_b})^2}\int\textrm{d}y \, \text{e}^{-iqy}\int\textrm{d}x_{1}\int\textrm{d}x_{2} \,\langle\sigma(x_1,y)\sigma(x_2,-y)\rangle_{ab}^\textrm{conn}  ,
\end{equation}
where $\langle\sigma\sigma\rangle_{ab}^\textrm{conn}$ denotes the connected correlator. We evaluate this correlator in the range (\ref{limits}) including also the first subleading corrections, and denote by $\hat{S}(q)$ the result that we obtain using this expression into (\ref{sf}) and performing the integral over $y$ from $-R/2$ to $R/2$. We find
\begin{equation}
\label{*8.16}
\hat{S}(q) = \frac{1}{mq^{2}} + \frac{c_{0}^{2}\sin Q}{m^{2}q} + \frac{2}{mR} \biggl[ 2\alpha_{2}^{2}\frac{\sin Q}{m^{2}q} + 2\alpha_{2}\frac{\cos Q}{mq^{2}} - \frac{\sin Q}{q^{3}} \biggr] + \mathcal{O}\left( R^{-2} \right)\,,
\end{equation}
with $Q=qR/2$. While $\alpha_2$ is a boundary coefficient, $c_0$ is specific of the bulk theory; it vanishes for the Ising universality class but takes a known non-zero value in other cases such as the three-state Potts universality class (see \cite{DV,DS2}). We stress that $\langle\sigma\sigma\rangle_{ab}^\textrm{conn}$ does not contain bulk correlations, so that (\ref{*8.16}) is entirely due to the interface. From (\ref{*8.16}) we have
\EQ
\lim_{R\to\infty}\hat{S}(q)=\frac{1}{mq^2}\,,\hspace{.6cm}q^2>0\,.
\label{Rinfinity}
\EN
This term is the one dominating at small $q$ in effective theories, where (in $d\geq 3$) it is obtained associating the long wavelength modes to free massless bosons with support on the plane corresponding to minimal interfacial area. On the other hand, the l.h.s. of (\ref{Rinfinity}) receives from (\ref{*8.16}) additional contributions at $q=0$; in particular, the term proportional to $c_0^2$ becomes $\frac{\pi c_0^2}{m^2}\,\delta(q)$ in the limit. These additional contributions reflect the specific form of the long range correlations which we exhibited above.

The paper is organized as follows. In the next section we recall the derivation of the order parameter profile as a warm up for the determination of the two-point function that we perform in section~3. Section~4 is devoted to the study of the interface structure factor, while section~5 contains some final remarks. Four appendices contain some developments of the analysis performed in the main body of the paper, as well as some mathematical aspects.

\section{One-point function}
\label{*sec.02}
\begin{figure}[htbp]
\centering
\includegraphics[width=12.2cm]{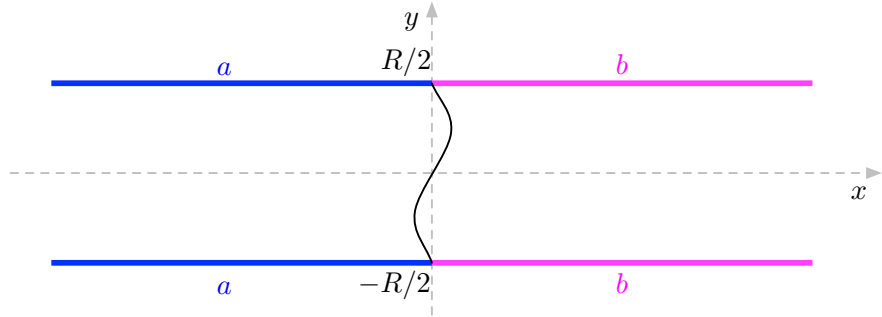}
\caption{The strip and boundary conditions considered throughout the paper, with a pictorial representation of the interface running between the boundary condition changing points.}
\label{fig.01}
\end{figure}

In this section we review the derivation of the order parameter one-point function \cite{DV,DS2} as an introduction to the calculation of the two-point function. As explained in the introduction, we consider a near critical system at phase coexistence in the strip geometry depicted in Fig.\ref{fig.01}, with boundary conditions on the edges favoring a phase $a$ for $x<0$ and a phase $b$ for $x>0$.
The fact that the system is close to criticality (i.e. to a point of second order phase transition) ensures that the bulk correlation length $\xi$ is much larger than microscopic scales and that all universal properties are described by a two-dimensional Euclidean field theory. The latter is related to a quantum field theory in one spatial dimension (with coordinate $x$) by analytic continuation to imaginary time, $y=it$. The fact that the system is at phase coexistence then means that the quantum theory possesses degenerate vacuum states $|\Omega_a\rangle$, one for each coexisting phase. In this (1+1)-dimensional case the elementary quantum excitations are kinks $K_{ab}(\theta)$ interpolating between two different vacua $\Omega_a$ and $\Omega_b$; the rapidity $\theta$ parameterizes energy and momentum of these relativistic particles as
\EQ
(e,p)=(m\cosh\theta,m\sinh\theta)\,,
\EN
where $m\propto 1/\xi$ is the kink mass. The trajctories of the kink $K_{ab}$ in imaginary time are domain walls separating phase $a$ from phase $b$.
The collection of all multikink states $|K_{a_1a_2}(\theta_1)K_{a_2a_3}(\theta_2)\\ \ldots K_{a_na_{n+1}}(\theta_n)\rangle$ form a complete basis. The boundary conditions on the edges of the strip play the role of boundary states for the imaginary time evolution, and can be expanded over the basis of kink states. For a boundary located at $y=it$ and boundary conditions changing from $a$ to $b$ at a spatial coordinate $x$ this expansion takes the form
\begin{equation}
\label{*1.01}
\vert B_{ab}(x;it) \rangle = \text{e}^{-itH+ixP} \biggl[ \int_{\mathbb{R}}\frac{\rd\theta}{2\pi} f_{ab}(\theta) \vert K_{ab}(\theta) \rangle + \dots \biggr] ,
\end{equation}
where $H$ and $P$ are the Hamiltonian and momentum operators of the one-dimensional quantum system, and the dots stay for multikink states interpolating between $\Omega_a$ and $\Omega_b$. As explained in the introduction, in this paper we restrict to universality classes for which the boundary conditions of Fig.\ref{fig.01} do not lead to the formation of a macroscopic layer of a third phase in the interfacial region (see \cite{DS2,localization} for a detailed analysis), and this ensures that $f_{ab}\neq 0$ in (\ref{*1.01}). At this point the partition function for the system reads
\begin{eqnarray} \nonumber
\label{*1.02}
\mathcal{Z}_{ab} & = & \langle B_{ab}(0;iR/2) \vert B_{ab}(0;-iR/2) \rangle \\
& \simeq & \int_{\mathbb{R}} \frac{\rd\theta}{2\pi} \vert f_{ab}(\theta) \vert^{2} \text{e}^{-mR\cosh\theta} \simeq \vert f_{ab}(0) \vert^{2} \frac{\text{e}^{-mR}}{\sqrt{2\pi m R}},
\label{partition}
\end{eqnarray}
where in the last line we took the limit for $mR$ large, which is needed for the emergence of phase separation and projects onto the lightest (single-kink) contribution in the expansion of the boundary states\footnote{We normalize the states according to $\langle K_{ab}(\theta_1)|K_{ab}(\theta_2)\rangle=2\pi\delta(\theta_1-\theta_2)$.}. Here and below the symbol $\simeq$ indicates omission of terms subleading in such a limit. It follows from (\ref{partition}) that the interfacial tension, corresponding to 
\EQ
-\lim_{R\to\infty}\frac{1}{R}\ln\mathcal{Z}_{ab}\,,
\EN
coincides with $m$. 

Along the same lines, the one-point function of the order parameter field $\sigma$ reads
\bea
\langle\sigma(x,y)\rangle_{ab} &=& \frac{1}{\mathcal{Z}_{ab}} \langle B_{ab}(0;iR/2) \vert \sigma(x,y)\vert B_{ab}(0;-iR/2) \rangle \nonumber \\
\label{*1.04}
& \simeq & \frac{1}{\mathcal{Z}_{ab}} \int_{\mathbb{R}^{2}}\frac{\rd\theta_{1}\rd\theta_{2}}{(2\pi)^{2}} \, \overline{f_{ab}}(\theta_{1}) f_{ab}(\theta_{2}) \mathcal{M}_{ab}^{\sigma}(\theta_{1}\vert\theta_{2}) \mathcal{U}_{x,y}^{+}(\theta_{1})\mathcal{U}_{_{x,y}}^{-}(\theta_{2})\,,
\eea
where we used the relation
\EQ
\sigma(x,y) = \text{e}^{ixP+yH} \sigma(0,0) \text{e}^{-ixP-yH}\,,
\label{shift}
\EN
and the notations
\EQ
\mathcal{U}_{x,y}^{\pm}(\theta) \equiv \text{e}^{-\left(\frac{mR}{2} \mp my\right)\cosh\theta \pm imx\sinh\theta}\,,
\EN
\EQ
\mathcal{M}_{ab}^{\sigma}(\theta_{1}\vert\theta_{2}) \equiv \langle K_{ba}(\theta_{1}) \vert \sigma(x,y) \vert K_{ab}(\theta_{2}) \rangle = \left\{
\begin{array}{ll}
\mathcal{F}_{ab}^{\sigma}(\theta_{12}) + 2\pi \langle\sigma\rangle_{a} \delta(\theta_{12}) \,, \quad \text{right}\,, \\
\\
\mathcal{F}_{ab}^{\sigma}(\theta_{12}) + 2\pi \langle\sigma\rangle_{b} \delta(\theta_{12}) \,, \quad \text{left}\,, \\
\end{array} 
\right.
\label{2particle}
\EN
\EQ
\theta_{12}\equiv\theta_1-\theta_2\,.
\label{theta12}
\EN
In the r.h.s. of (\ref{2particle}) we made explicit the decomposition of the matrix element of the field into a connected and a disconnected part, the latter corresponding to particle annihilation. Pictorially we have
\begin{equation}
\label{*1.05}
\hspace{-5mm}\mathcal{M}_{ab}^{\sigma}(\theta_{1}\vert\theta_{2})=\quad \newfig{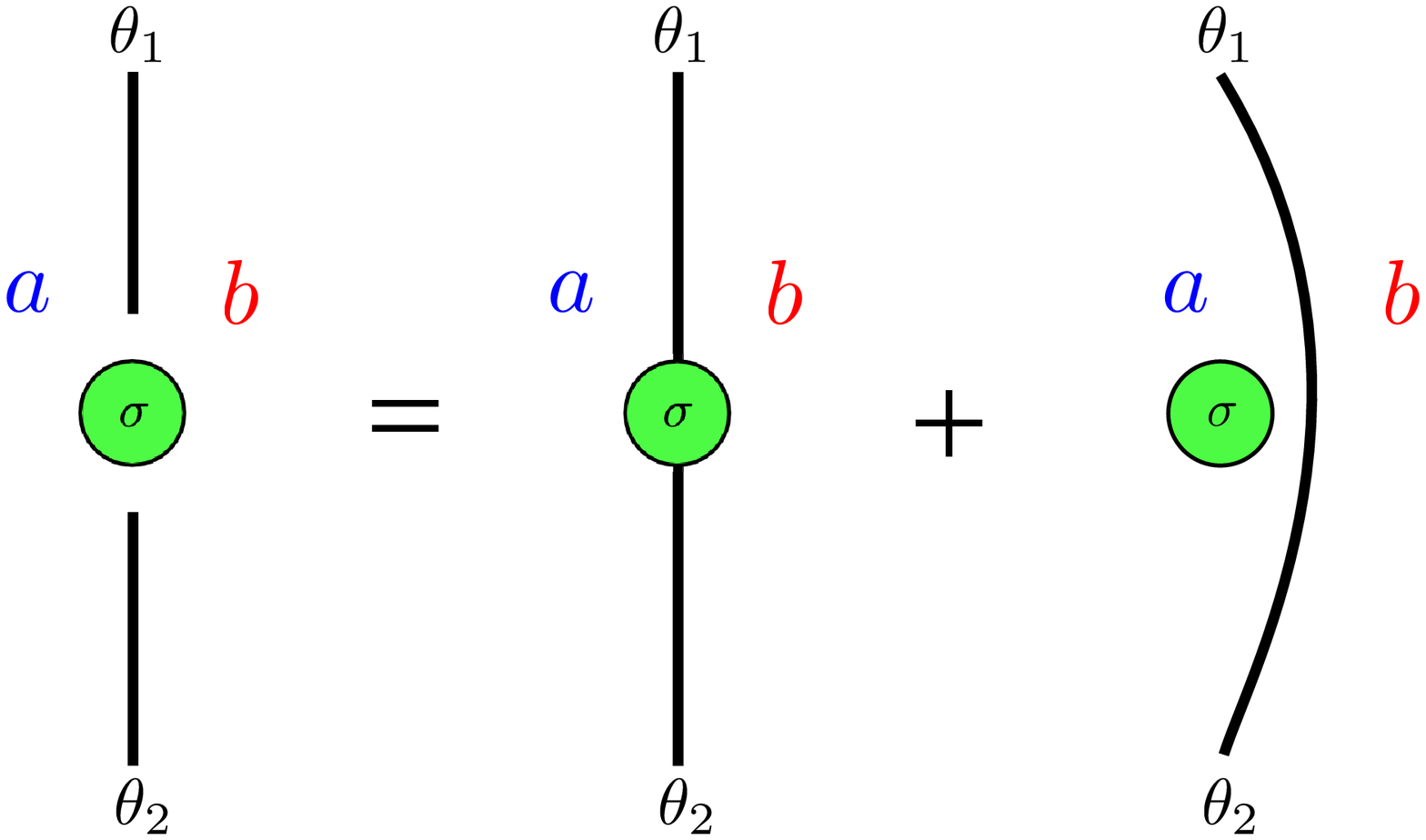} \hspace{50mm} 
\end{equation}
for the case in which the annihilation takes place to the right of the field in the Euclidean plane. The right-left alternative is ultimately responsible for the presence of the kinematical (or annihilation pole) \cite{Smirnov,DC,vortex}
\begin{equation}
\label{*1.06}
\mathcal{F}_{ab}^{\sigma}(\theta_{12}) \simeq \frac{i\Delta\langle\sigma\rangle}{\theta_{12}}\, , \hspace{5mm}\theta_1\to\theta_2\,;\hspace{1cm} \Delta\langle\sigma\rangle \equiv \langle\sigma\rangle_{a} - \langle\sigma\rangle_{b}\,.
\end{equation}
Plugging (\ref{*1.06}) into (\ref{*1.04}) we can write\footnote{We disregard the additive constant contributed by the disconneted part of the matrix element; it is associated to the regularization of the pole and drops out when taking the derivative (\ref{*1.08}).}
\begin{equation}
\label{*1.07}
\langle\sigma(x,y)\rangle_{ab} \simeq \frac{i\Delta\langle\sigma\rangle}{2\pi^{\frac{3}{2}}} \int_{\mathbb{R}^{2}} \frac{\rd\theta_{1}\rd\theta_{2}}{\theta_{12}} \, \mathcal{U}_{\eta,\epsilon}^{+}(\theta_{1})\mathcal{U}_{\eta,\epsilon}^{-}(\theta_{2})\,,
\end{equation}
where we introduced the notations $\mathcal{U}_{\eta,\epsilon}^{\pm}(\theta) = \text{e}^{-\frac{1 \mp \epsilon}{2}\theta^{2}\pm i\eta\theta}$ and 
\EQ
\eta = \frac{x}{\lambda}\,,\hspace{1cm}\epsilon=\frac{2y}{R}\,,\hspace{1cm}
\lambda=\sqrt{\frac{R}{2m}}\,.
\label{etaeps}
\EN
The pole in (\ref{*1.07}) should be intended in the regularized form $\theta^{-1} = \mathcal{P}\left(\theta^{-1}\right) \pm \pi i \delta(\theta)$, and can be easily handled taking the derivative
\begin{equation}
\label{*1.08}
\partial_{\eta}\langle\sigma(x,y)\rangle_{ab}\simeq -\frac{\Delta\langle\sigma\rangle}{2\pi^{\frac{3}{2}}} \int_{\mathbb{R}^{2}} \rd\theta_{1}\rd\theta_{2} \, \mathcal{U}_{\eta,\epsilon}^{+}(\theta_{1})\mathcal{U}_{\eta,\epsilon}^{-}(\theta_{2})= -\frac{\Delta\langle\sigma\rangle}{\sqrt{\pi}\kappa} \text{e}^{-\chi^{2}} ,
\end{equation}
where we introduced the additional notations
\EQ
\chi=\frac{\eta}{\kappa}\,,\hspace{1cm}\kappa=\sqrt{1-\epsilon^{2}}\,.
\label{chikappa}
\EN
Integrating back in $\eta$ with the boundary condition $\lim_{x\rightarrow+\infty}\langle\sigma(x,y)\rangle_{ab} = \langle\sigma\rangle_{b}$ we finally obtain
\begin{equation}
\label{*1.10}
\langle\sigma(x,y)\rangle_{ab}\simeq \frac{\langle\sigma\rangle_{a}+\langle\sigma\rangle_{b}}{2}-\frac{\langle\sigma\rangle_{a}-\langle\sigma\rangle_{b}}{2}\,\text{erf}(\chi) .
\end{equation}

\begin{figure}[t]
\centering
\includegraphics[width=8.5cm]{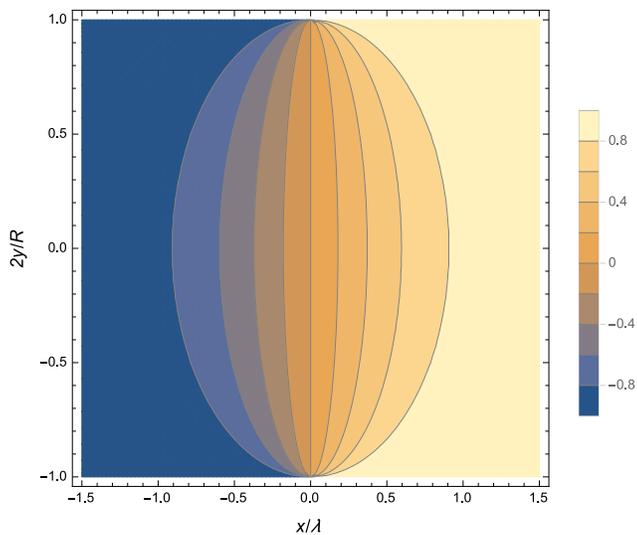}
\caption{The order parameter profile (\ref{*1.10}).}
\label{fig.02}
\end{figure}

It is easy to see that this leading contribution to the order parameter profile, which is entirely due to the pole term (\ref{*1.06}) and correctly interpolates between $\langle\sigma\rangle_a$ at $x=-\infty$ and $\langle\sigma\rangle_b$ at $x=+\infty$ (Fig.\ref{fig.02}), amounts to the presence of a fluctuating interface whose configurations sharply separate two pure phases. Indeed, denoting by $P_{1}(x;y)\rd x$ the probability that such an interface intersects the line of ordinate $y$ in the infinitesimal interval $(x,x+\rd x)$, the corresponding profile reads
\begin{equation}
\label{*1.11}
\langle\sigma(x,y)\rangle_{ab}^\textrm{sharp} = \langle\sigma\rangle_{a} \int_{x}^{+\infty}\rd u \, P_{1}(u;y) + \langle\sigma\rangle_{b} \int_{-\infty}^{x}\rd u \, P_{1}(u;y)\,.
\end{equation}
The derivative with respect to $x$ matches (\ref{*1.08}) for a passage probability density 
\begin{equation}
\label{*1.12}
P_{1}(x;y) = \frac{\text{e}^{-\chi^{2}}}{\sqrt{\pi}\kappa\lambda}
\end{equation}
which correctly satisfies $\int_{-\infty}^{\infty}\textrm{d}x\,P_1(x;y)=1$ and is plotted in Fig.\ref{fig.03}. As we explain in appendix~\ref{App_B}, the result (\ref{*1.12}) shows that the interface behaves as a Brownian bridge connecting the boundary condition changing points on the edges of the strip. 
It can also be shown that subleading corrections to (\ref{*1.10}) in the large $R$ expansion account for the internal structure of the interface (see \cite{DV,DS2} and appendix~\ref{app_correlator_prob} below). 

\begin{figure}[t]
\centering
\includegraphics[width=8.5cm]{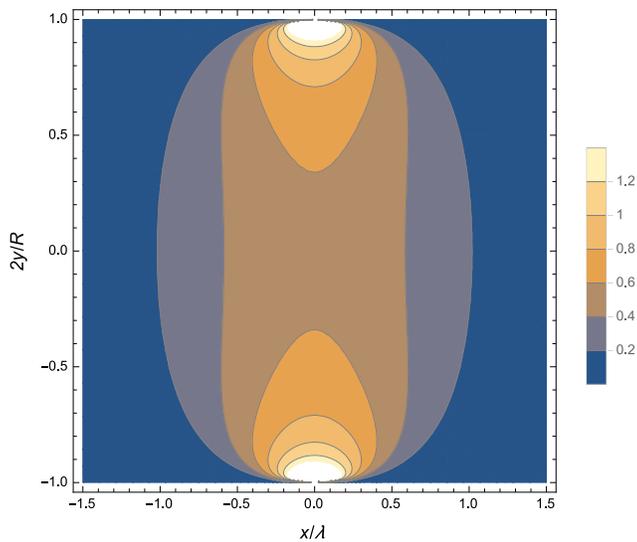}
\caption{The rescaled passage probability density $\lambda^{-1}P_{1}(x;y)$.}
\label{fig.03}
\end{figure}

\section{Two-point function}
\subsection{Field theoretical derivation}
\label{*sec.03}
The formalism of the previous section can now be used for the determination of the two-point function\footnote{We perform the computation for the general case of correlations between two different components $\sigma_1$ and $\sigma_2$ of the order parameter field.} 
\begin{equation}
\label{*3.01}
\langle\sigma_{1}(x_{1},y_{1})\sigma_{2}(x_{2},y_{2})\rangle_{ab} = \frac{1}{\mathcal{Z}_{ab}}\langle B_{ab}(0;iR/2) \vert \sigma_{1}(x_{1},y_{1})\sigma_{2}(x_{2},y_{2}) \vert B_{ab}(0;-iR/2) \rangle
\end{equation}
in the limits in which $R\gg y_1-y_2\gg\xi$, and the distance of $y_1$ and $y_2$ from the edges of the strip is also much larger than $\xi$. This ensures that, upon expansion of the boundary states and insertion of a complete set of multikink states in between the two fields, the single-kink state gives the dominant contribution, so that 
\begin{eqnarray}\nonumber
\label{*3.02}
\langle\sigma_{1}(x_{1},y_{1})\sigma_{2}(x_{2},y_{2})\rangle_{ab} & \simeq & \frac{1}{\mathcal{Z}_{ab}} \int_{\mathbb{R}^{3}} \frac{\rd\theta_{1}\rd\theta_{2}\rd\theta_{3}}{(2\pi)^{3}} \, \overline{f_{ab}}(\theta_{1})f_{ab}(\theta_{2}) \, \langle K_{ba}(\theta_{1}) \vert \sigma_{1}(x_{1},y_{1}) \vert K_{ab}(\theta_{3})\rangle \\
& \times & \langle K_{ba}(\theta_{3}) \vert \sigma_{2}(x_{2},y_{2}) \vert K_{ab}(\theta_{2}) \rangle \text{e}^{-\frac{mR}{2}\left(\cosh\theta_{1}+\cosh\theta_{2}\right)} .
\end{eqnarray}
Using (\ref{shift}) and defining
\begin{eqnarray} \nonumber
\label{*3.03}
\ln\mathcal{E}(\theta_{1},\theta_{2},\theta_{3}) & = & -\left(\frac{mR}{2}-my_{1}\right)\cosh\theta_{1} - \left(\frac{mR}{2}+my_{2}\right)\cosh\theta_{2} - \left(my_{1}-my_{2}\right)\cosh\theta_{3} \\
& + & imx_{1}\left(\sinh\theta_{1}-\sinh\theta_{3}\right)+ imx_{2}\left(\sinh\theta_{3}-\sinh\theta_{2}\right) ,
\end{eqnarray}
we can write
\begin{equation}
\label{*3.04}
\langle\sigma_{1}(x_{1},y_{1})\sigma_{2}(x_{2},y_{2})\rangle_{ab}\simeq \frac{1}{\mathcal{Z}_{ab}} \int_{\mathbb{R}^{3}} \frac{\rd\theta_{1}\rd\theta_{2}\rd\theta_{3}}{(2\pi)^{3}} \, \overline{f_{ab}}(\theta_{1})f_{ab}(\theta_{2}) \, \mathcal{M}_{ab}^{\sigma_{1}}(\theta_{1}\vert\theta_{3}) \mathcal{M}_{ab}^{\sigma_{2}}(\theta_{3}\vert\theta_{2}) \, \mathcal{E}(\theta_{1},\theta_{2},\theta_{3}) , \\
\end{equation}
then, since small rapidities dominate in the limits we consider,
\begin{eqnarray}
\label{*3.05}
\langle\sigma_{1}(x_{1},y_{1})\sigma_{2}(x_{2},y_{2})\rangle_{ab} & \simeq & \frac{|f_{ab}(0)|^2}{\mathcal{Z}_{ab}} \int_{\mathbb{R}^{3}} \frac{\rd\theta_{1}\rd\theta_{2}\rd\theta_{3}}{(2\pi)^{3}} \, \mathcal{M}_{ab}^{\sigma_{1}}(\theta_{1}\vert\theta_{3}) \mathcal{M}_{ab}^{\sigma_{2}}(\theta_{3}\vert\theta_{2}) \widetilde{\mathcal{E}}(\theta_{1},\theta_{2},\theta_{3}) ,
\end{eqnarray}
with
\begin{equation}
\label{*3.06}
\ln\widetilde{\mathcal{E}}(\theta_{1},\theta_{2},\theta_{3}) = -mR - \frac{mR}{4} \biggl[ \left(1-\epsilon_{1}\right)\theta_{1}^{2} + \left(1+\epsilon_{2}\right)\theta_{2}^{2}+\left(\epsilon_{1} - \epsilon_{2}\right)\theta_{3}^{2} \biggr] + imx_{1}\theta_{13} + imx_{2}\theta_{32} ,
\end{equation}
and $\epsilon_{j}=\frac{2y_{j}}{R}$. The matrix elements can be decomposed as in (\ref{2particle}); pictorially
\begin{equation}
\label{*3.07}
\hspace{-5mm}\mathcal{M}_{ab}^{\sigma_{1}}(\theta_{1}\vert\theta_{3}) \mathcal{M}_{ab}^{\sigma_{2}}(\theta_{3}\vert\theta_{2})= \quad \newfig{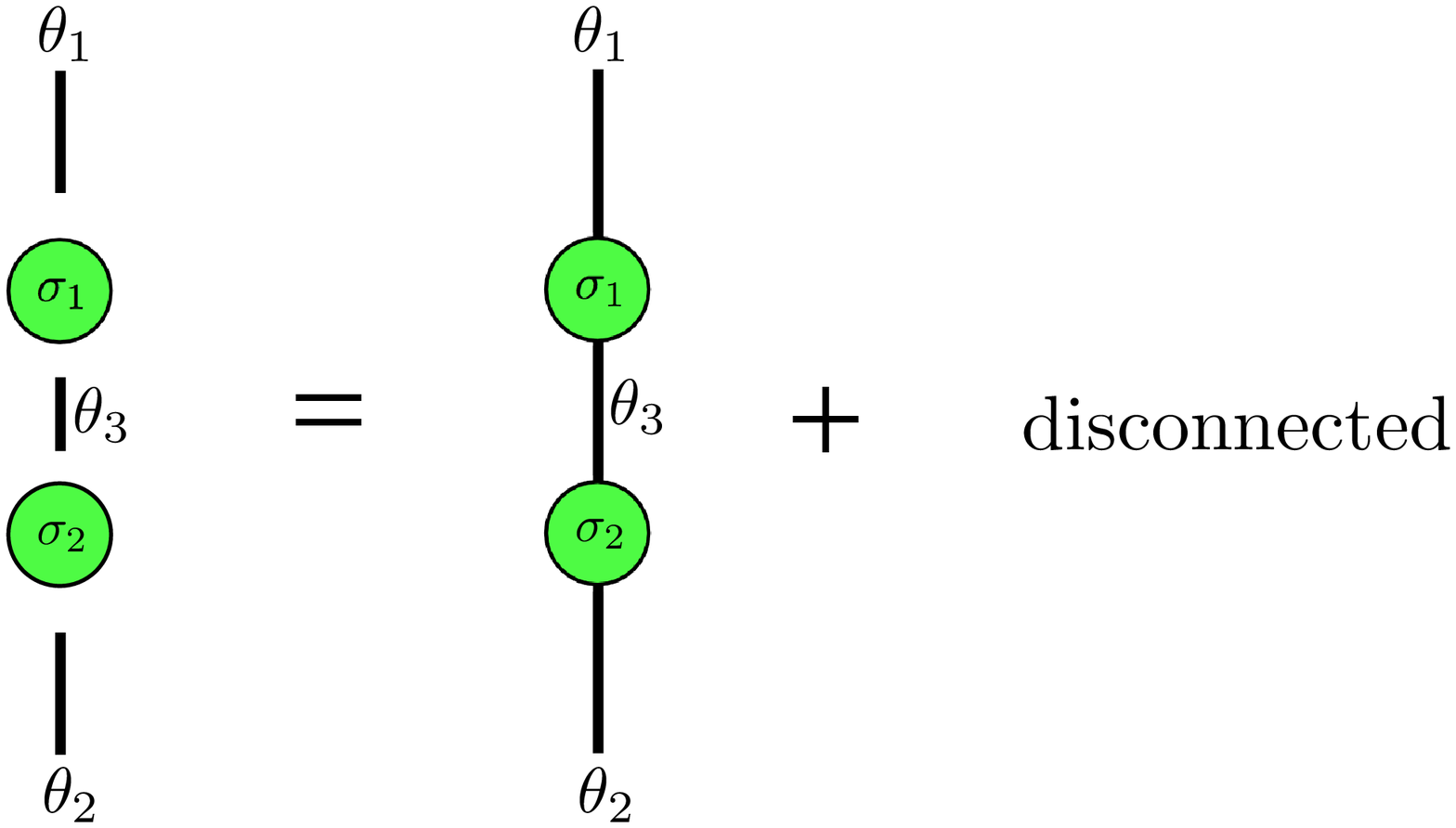} \hspace{41mm}.
\end{equation}
Let us consider first the contribution coming from the connected parts of the matrix elements (we will denote it by a superscript CP); the leading contribution comes from the pole (\ref{*1.06}), and reads
\EQ
\label{*3.08}
\langle\sigma(x_{1},y_{1})\sigma(x_{2},y_{2})\rangle_{ab}^{\text{CP}}\simeq \frac{|f_{ab}(0)|^2}{\mathcal{Z}_{ab}} \int_{\mathbb{R}^{3}}\frac{\rd\theta_{1}\rd\theta_{2}\rd\theta_{3}}{(2\pi)^{3}} \biggl[ \frac{i^{2}\Delta\langle\sigma_{1}\rangle\Delta\langle\sigma_{2}\rangle}{\theta_{13}\theta_{32}} \biggr] \widetilde{\mathcal{E}}(\theta_{1},\theta_{2},\theta_{3}) ,
\EN
where $\Delta\langle\sigma_{j}\rangle\equiv\langle\sigma_{j}\rangle_{a}-\langle\sigma_{j}\rangle_{b}$. It is convenient to define
\begin{equation}
\label{*3.09}
\langle\sigma(x_{1},y_{1})\sigma(x_{2},y_{2})\rangle_{ab}^{\text{CP}}\simeq \frac{\Delta\langle\sigma_{1}\rangle\Delta\langle\sigma_{2}\rangle}{4}\,\mathcal{G}(\eta_{1},\epsilon_{1};\eta_{2},\epsilon_{2})\,,
\end{equation}
where $\eta_{j}=x_{j}/\lambda$, $\epsilon_{j}=2y_{j}/R$ and
\EQ
\label{*3.10}
\mathcal{G}(\eta_{1},\epsilon_{1};\eta_{2},\epsilon_{2}) = \frac{1}{\pi^{5/2}}\int_{\mathbb{R}^{3}} \frac{\rd\theta_{1}\rd\theta_{2}\rd\theta_{3}}{\theta_{13}{\theta_{23}}} \text{e}^{-\frac{1-\epsilon_{1}}{2}\theta_{1}^{2}-\frac{1+\epsilon_{2}}{2}\theta_{2}^{2}-\frac{\epsilon_{1}-\epsilon_{2}}{2}\theta_{3}^{2}+i\eta_{1}\theta_{13}+i\eta_{2}\theta_{32}} .
\EN
The explicit computation of the function (\ref{*3.10}) will be performed later in this section. For the time being we give a simplified integral representation that will be useful in the coming sections. Calculations are simplified if we apply the differential operator $\partial_{\eta_{1},\eta_{2}}^{2}$, which removes the poles. This leaves us with Gaussian integrals, and integrating over $\theta_{1}$ and $\theta_{2}$ we obtain
\EQ
\label{*3.11}
\partial_{\eta_{1},\eta_{2}}^{2}\mathcal{G}(\eta_{1},\epsilon_{1},\eta_{2},\epsilon_{2}) = \frac{2}{\pi^{3/2}}\frac{\text{e}^{-\frac{\eta_{1}^{2}}{2(1-\epsilon_{1})}-\frac{\eta_{2}^{2}}{2(1+\epsilon_{2})}}}{\sqrt{(1-\epsilon_{1})(1+\epsilon_{2})}}  \int_{\mathbb{R}} \rd\theta \, \text{e}^{-\frac{\epsilon_{1}-\epsilon_{2}}{2}\theta^{2}+i(\eta_{2}-\eta_{1})\theta}\,;
\EN
integrating back\footnote{This operation produces integration constants, but it is simple to show that they can be reabsorbed in the contribution of the disconnected parts of the matrix elements. Hence, we set these constants to zero in (\ref{*3.12}).} over $\eta_{1}$ and $\eta_{2}$ and using the identity (\ref{*A.05}) we express $\mathcal{G}$ through the single-integral representation
\EQ
\label{*3.12}
\mathcal{G}(\eta_{1},\epsilon_{1};\eta_{2},\epsilon_{2}) = \frac{1}{\sqrt{\pi}} \int_{\mathbb{R}} \rd\theta \, \text{e}^{-\theta^{2}} \text{erf}\left(\frac{\eta_{1}+i(1-\epsilon_{1})\theta}{\sqrt{2(1-\epsilon_{1})}}\right) \text{erf}\left(\frac{\eta_{2}-i(1+\epsilon_{2})\theta}{\sqrt{2(1+\epsilon_{2})}}\right) .
\EN

Let us now consider the contributions coming from the disconnected parts in (\ref{*3.07}). Pictorially, these disconnected parts correspond to
\EQ \nonumber
\mathcal{D}_{1_{L}} = \newfig{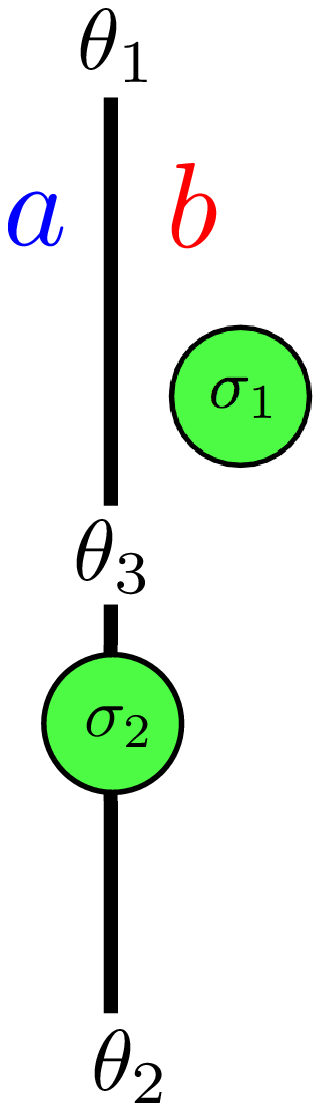} \quad,
\quad \mathcal{D}_{1_{R}} = \newfig{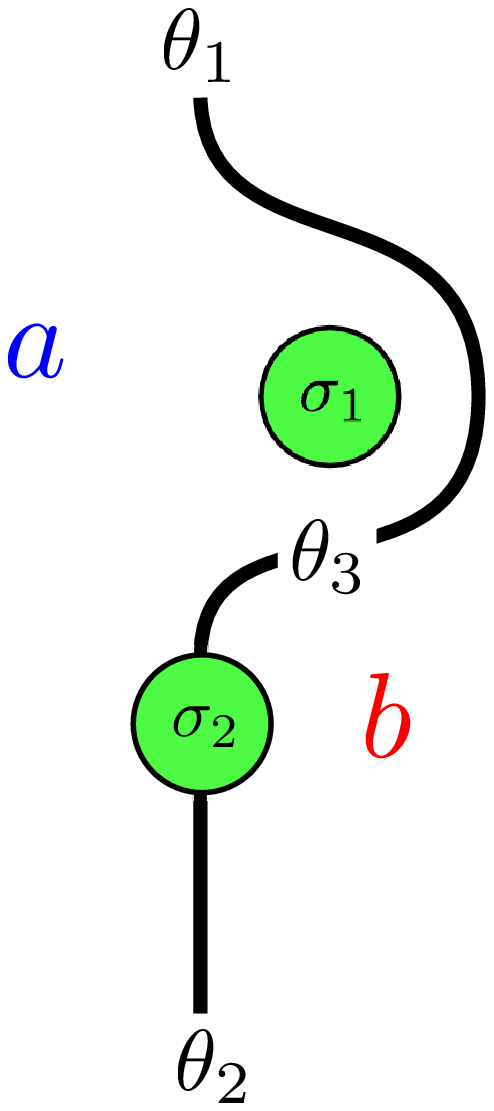},
\quad \mathcal{D}_{2_{L}} = \newfig{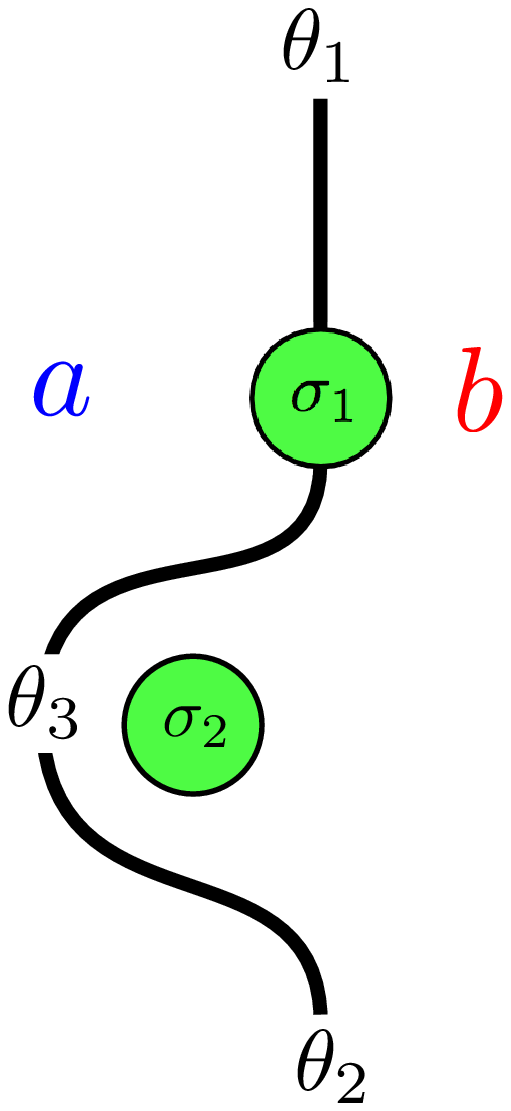} ,
\, \quad \mathcal{D}_{2_{R}} = \newfig{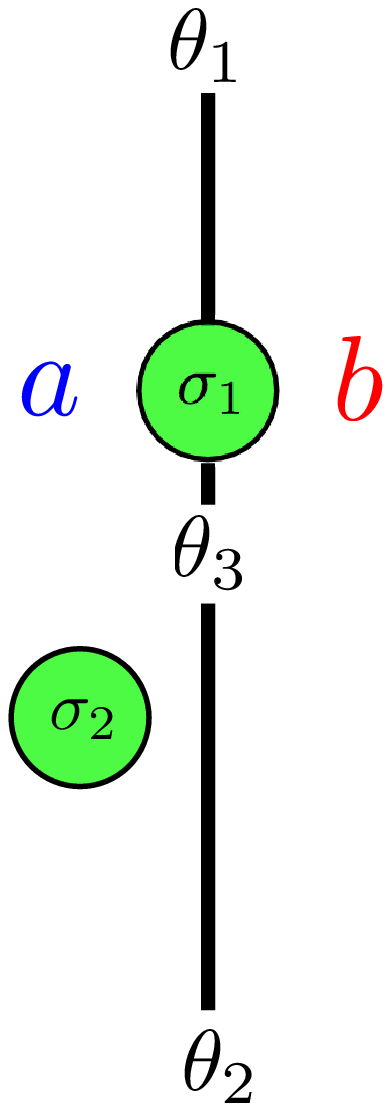}, \\
\EN
\EQ \nonumber
\mathcal{D}_{1_{L}2_{L}} = \newfig{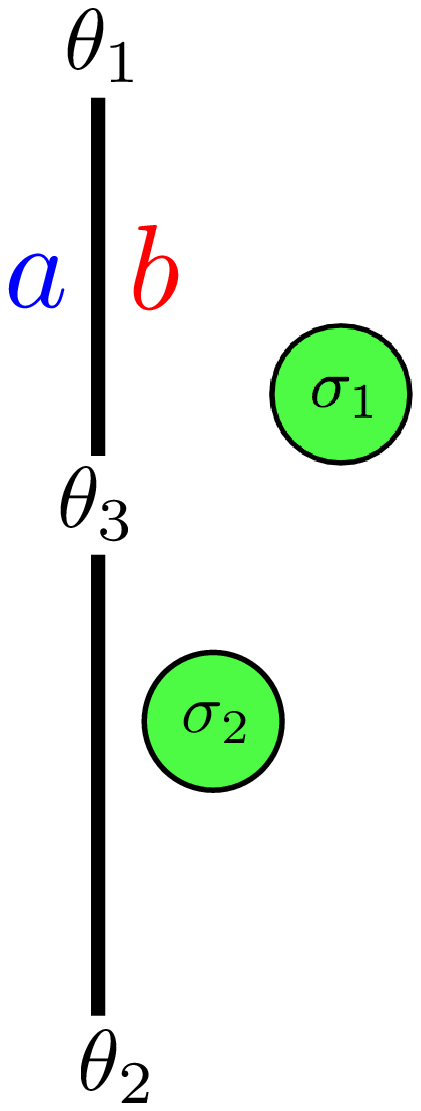} \quad,
\,\, \mathcal{D}_{1_{L}2_{R}} = \newfig{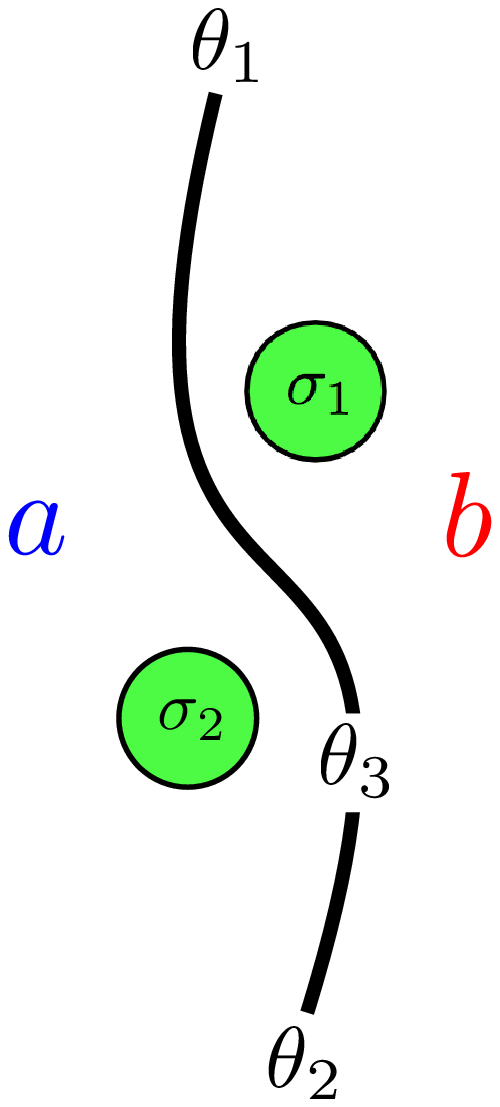} \quad ,
\quad \mathcal{D}_{1_{R}2_{L}} = \newfig{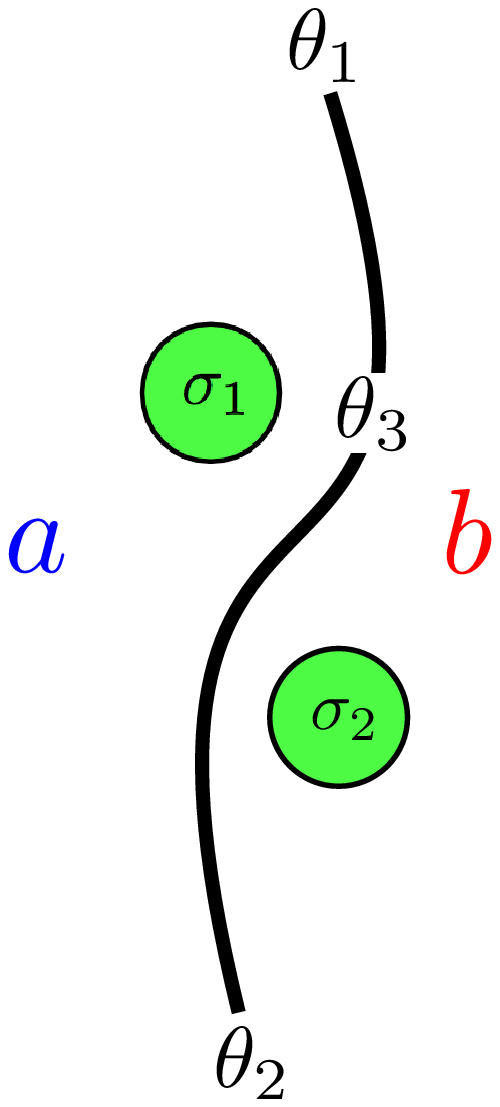},
\,\, \mathcal{D}_{1_{R}2_{R}} = \newfig{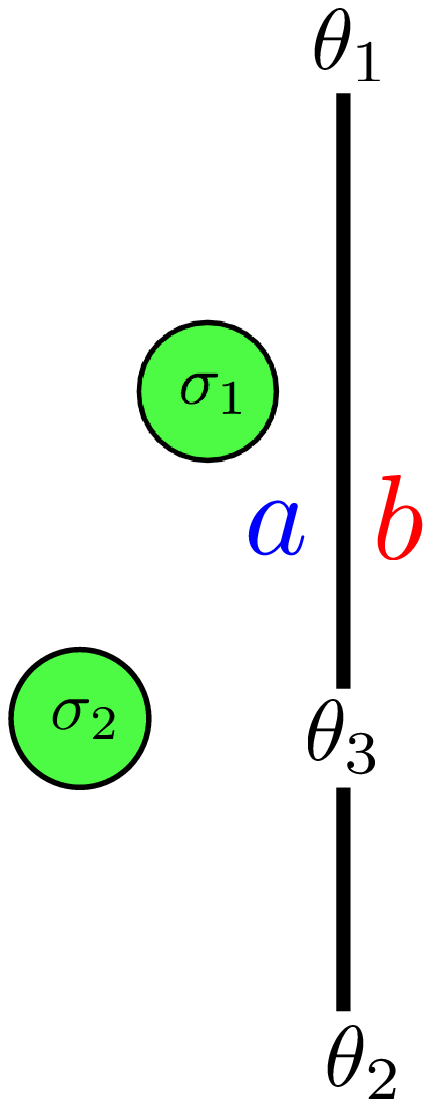} ,
\EN
and give the following contributions to the two-point function
\begin{eqnarray} \nonumber
\label{*3.14}
\mathcal{D}_{1_{L}} = 2\pi i \langle\sigma_{1}\rangle_{b} \Delta\langle\sigma_{2}\rangle \frac{\delta(\theta_{13})}{\theta_{32}} & \longrightarrow & -\langle\sigma_{1}\rangle_{b}\Delta\langle\sigma_{2}\rangle \text{erf}(\chi_{2}) , \\ \nonumber
\mathcal{D}_{1_{R}} = 2\pi i \langle\sigma_{1}\rangle_{a} \Delta\langle\sigma_{2}\rangle \frac{\delta(\theta_{13})}{\theta_{32}} & \longrightarrow & -\langle\sigma_{1}\rangle_{a}\Delta\langle\sigma_{2}\rangle \text{erf}(\chi_{2}) , \\ \nonumber
\mathcal{D}_{2_{L}} = 2\pi i \langle\sigma_{2}\rangle_{b} \Delta\langle\sigma_{1}\rangle \frac{\delta(\theta_{32})}{\theta_{13}} & \longrightarrow & -\langle\sigma_{2}\rangle_{b}\Delta\langle\sigma_{1}\rangle \text{erf}(\chi_{1}) , \\
\mathcal{D}_{2_{R}} = 2\pi i \langle\sigma_{2}\rangle_{a} \Delta\langle\sigma_{1}\rangle \frac{\delta(\theta_{32})}{\theta_{13}} & \longrightarrow & -\langle\sigma_{2}\rangle_{a}\Delta\langle\sigma_{1}\rangle \text{erf}(\chi_{1}),\nonumber
\end{eqnarray}
\begin{eqnarray} \nonumber
\label{*3.15}
\mathcal{D}_{1_{L}2_{L}} = (2\pi)^{2} \langle\sigma_{1}\rangle_{b}\langle\sigma_{2}\rangle_{b} \delta(\theta_{13})\delta(\theta_{23}) & \longrightarrow & \langle\sigma_{1}\rangle_{b}\langle\sigma_{2}\rangle_{b} , \\ \nonumber
\mathcal{D}_{1_{L}2_{R}} = (2\pi)^{2} \langle\sigma_{1}\rangle_{b}\langle\sigma_{2}\rangle_{a} \delta(\theta_{13})\delta(\theta_{23}) & \longrightarrow & \langle\sigma_{1}\rangle_{b}\langle\sigma_{2}\rangle_{a} , \\ \nonumber
\mathcal{D}_{1_{R}2_{L}} = (2\pi)^{2} \langle\sigma_{1}\rangle_{a}\langle\sigma_{2}\rangle_{b} \delta(\theta_{13})\delta(\theta_{23}) & \longrightarrow & \langle\sigma_{1}\rangle_{a}\langle\sigma_{2}\rangle_{b} , \\
\mathcal{D}_{1_{R}2_{R}} = (2\pi)^{2} \langle\sigma_{1}\rangle_{a}\langle\sigma_{2}\rangle_{a} \delta(\theta_{13})\delta(\theta_{23}) & \longrightarrow & \langle\sigma_{1}\rangle_{a}\langle\sigma_{2}\rangle_{a} .\nonumber
\end{eqnarray}
The prescription is to take the arithmetic average of passage left and right \cite{DS2}, so that, putting all together, we finally obtain
\begin{eqnarray} \nonumber
\label{*3.16}
\langle\sigma_{1}(x_{1},y_{1})\sigma_{2}(x_{2},y_{2})\rangle_{ab} &\simeq & \frac{\Delta\langle\sigma_{1}\rangle\Delta\langle\sigma_{2}\rangle}{4} \, \mathcal{G}(\eta_{1},\epsilon_{1};\eta_{2},\epsilon_{2}) - \widetilde{\langle\sigma_{2}\rangle}\frac{\Delta\langle\sigma_{1}\rangle}{2}\text{erf}(\chi_{1}) + \\
& - & \widetilde{\langle\sigma_{1}\rangle}\frac{\Delta\langle\sigma_{2}\rangle}{2}\text{erf}(\chi_{2}) + \widetilde{\langle\sigma_{1}\rangle}\widetilde{\langle\sigma_{2}\rangle} ,
\end{eqnarray}
where we used the notation 
\EQ
\widetilde{\langle\sigma_{j}\rangle} = \frac{\langle\sigma_{j}\rangle_{a}+\langle\sigma_{j}\rangle_{b}}{2}\,.
\label{average}
\EN
For $\sigma_1=\sigma_2=\sigma$ (\ref{*3.16}) becomes
\begin{eqnarray} \nonumber
\label{*3.17}
\langle\sigma(x_{1},y_{1})\sigma(x_{2},y_{2})\rangle_{ab} & = & \frac{\left(\langle\sigma\rangle_{a}-\langle\sigma\rangle_{b}\right)^{2}}{4}\mathcal{G}(\eta_{1},\epsilon_{1};\eta_{2},\epsilon_{2}) + \frac{\langle\sigma\rangle_{b}^{2}-\langle\sigma\rangle_{a}^{2}}{4} \bigl[ \text{erf}(\chi_{1}) + \text{erf}(\chi_{2}) \bigr] + \\
& + & \frac{\left(\langle\sigma\rangle_{a}+\langle\sigma\rangle_{b}\right)^{2}}{4} .
\end{eqnarray}

\begin{figure}[t]
\centering
\includegraphics[width=10cm]{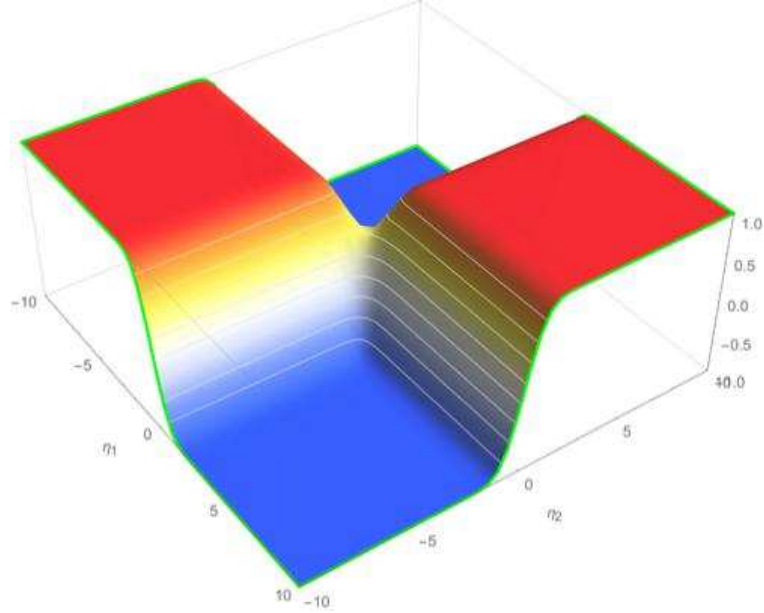}
\caption{The scaling function $\mathcal{G}(\eta_{1},\epsilon;\eta_{2},-\epsilon;)$ for $\epsilon=0.3$. 
}
\label{fig.G}
\end{figure}

Notice that, using (\ref{*3.12}) and the integral $I$ of appendix \ref{App_A}, one obtains
\begin{eqnarray}
\label{*3.18}
\lim_{\eta_{2}\rightarrow \pm \infty} \mathcal{G}(\eta_{1},\epsilon_{1};\eta_{2},\epsilon_{2}) & = &  \pm \text{erf}(\chi_{1}) , \\
\label{*3.19}
\lim_{\eta_{1}\rightarrow \pm \infty} \mathcal{G}(\eta_{1},\epsilon_{1};\eta_{2},\epsilon_{2}) & = &  \pm \text{erf}(\chi_{2}) ,
\end{eqnarray}
and then the cluster properties
\begin{eqnarray} \nonumber
\label{*3.20}
\lim_{x_{1}\rightarrow -\infty} \langle\sigma_{1}(x_{1},y_{1})\sigma_{2}(x_{2},y_{2})\rangle_{ab} & = & \langle\sigma_{1}\rangle_{a} \langle\sigma_{2}(x_{2},y_{2})\rangle_{ab} , \\ \nonumber
\lim_{x_{1}\rightarrow +\infty} \langle\sigma_{1}(x_{1},y_{1})\sigma_{2}(x_{2},y_{2})\rangle_{ab} & = & \langle\sigma_{1}\rangle_{b} \langle\sigma_{2}(x_{2},y_{2})\rangle_{ab} , \\ \nonumber
\lim_{x_{2}\rightarrow -\infty} \langle\sigma_{1}(x_{1},y_{1})\sigma_{2}(x_{2},y_{2})\rangle_{ab} & = & \langle\sigma_{2}\rangle_{a} \langle\sigma_{1}(x_{1},y_{1})\rangle_{ab} , \\
\lim_{x_{2}\rightarrow +\infty} \langle\sigma_{1}(x_{1},y_{1})\sigma_{2}(x_{2},y_{2})\rangle_{ab} & = & \langle\sigma_{2}\rangle_{b} \langle\sigma_{1}(x_{1},y_{1})\rangle_{ab} ,
\end{eqnarray}
in terms of the one-point functions computed in the previous section. 

The function (\ref{*3.12}) can be expressed in a closed form thanks to the integral $G$ discussed in appendix \ref{App_A}, which allows us to write $\mathcal{G}(\eta_{1},\epsilon_{1};\eta_{2},\epsilon_{2}) = G(i \sqrt{\frac{1-\epsilon_{1}}{2}},\frac{\eta_{1}}{\sqrt{2(1-\epsilon_{1})}},-i \sqrt{\frac{1+\epsilon_{2}}{2}},\frac{\eta_{2}}{\sqrt{2(1+\epsilon_{2})}})$, and 
\begin{equation}
\label{*3.21}
\mathcal{G}(\eta_{1},\epsilon_{1};\eta_{2},\epsilon_{2}) = \text{sign}(\eta_{1}\eta_{2}) - 4T(\sqrt{2}\chi_{1},Q_{1})- 4T(\sqrt{2}\chi_{2},Q_{2}),\hspace{.5cm}\eta_{1},\eta_{2}\neq 0\,,
\end{equation}
where $\chi_{j}=\frac{\eta_{j}}{\sqrt{1-\epsilon_{j}^{2}}}$, $T$ is Owen's $T$ function and
\begin{eqnarray} \nonumber
\label{*3.22}
Q_{1} & = & \sqrt{\frac{(1-\epsilon_{1})(1+\epsilon_{2})}{2(\epsilon_{1}-\epsilon_{2})}} \left( \frac{\eta_{2}}{\eta_{1}} \frac{1+\epsilon_{1}}{1+\epsilon_{2}}-1 \right) , \\
Q_{2} & = & \sqrt{\frac{(1-\epsilon_{1})(1+\epsilon_{2})}{2(\epsilon_{1}-\epsilon_{2})}} \left( \frac{\eta_{1}}{\eta_{2}} \frac{1-\epsilon_{2}}{1-\epsilon_{1}}-1 \right) ;\nonumber
\end{eqnarray}
if at least one of the two fields, say $\sigma_{2}$, is placed along the vertical axis one uses instead the representation
\begin{equation}
\label{*3.23}
\mathcal{G}(\eta_{1},\epsilon_{1};0,\epsilon_{2}) = 4 T\left(\sqrt{2}\chi_{1},\sqrt{\frac{(1-\epsilon_{1})(1+\epsilon_{2})}{2(\epsilon_{1}-\epsilon_{2})}}\right) .
\end{equation}
The passage from (\ref{*3.21}) to (\ref{*3.23}) is smooth and follows from the properties of the function $T$ collected in appendix A. A plot of the function (\ref{*3.21}) is shown in Fig.\ref{fig.G}. 
The result (\ref{leading}) follows from (\ref{*3.17}) and (\ref{*3.21}).

\subsection{Probabilistic interpretation}
\label{*sec.04}
\begin{figure}[t]
\centering
\includegraphics[width=8.2cm]{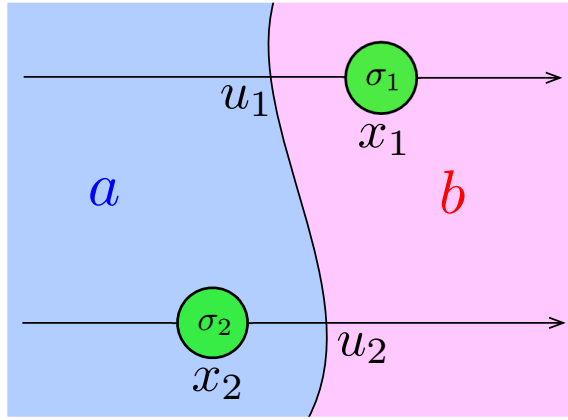}
\caption{The sharply separated phases for $u_{1}<x_{1}$ and $u_{2}>x_{2}$. For this configuration $\Gamma_{ab}=\langle\sigma_{1}\rangle_{b}\langle\sigma_{2}\rangle_{a}$.}
\label{fig.04}
\end{figure}

We now show that, similarly to what we saw for the one-point function, also the results (\ref{*3.17}), (\ref{*3.21}) can be interpreted in terms of a fluctuating interface whose configurations sharply separate two pure phases. Indeed, within this picture we now write
\begin{equation}
\label{*4.01}
\langle\sigma_{1}(x_{1},y_{1})\sigma_{2}(x_{2},y_{2})\rangle_{ab}^\textrm{sharp} = \int_{\mathbb{R}^{2}}\rd u_{1} \rd u_{2} \, P_{2}(u_{1},y_{1};u_{2},y_{2}) \, \Gamma_{ab}(x_{1},y_{1};x_{2},y_{2}\vert u_{1},u_{2})\,,
\end{equation}
where $P_{2}(u_{1},y_{1};u_{2},y_{2}) \textrm{d}u_{1} \textrm{d}u_{2}$ is the probability that the interface intersects the line of ordinate $y_1$ in the interval $(u_1,u_1+\textrm{d}u)$ {\it and} the line of ordinate $y_2$ in the interval $(u_2,u_2+\textrm{d}u)$; $\Gamma_{ab}(x_{1},y_{1};x_{2},y_{2}\vert u_{1},u_{2})$ is the value of $\sigma_1(x_{1},y_{1})\sigma_{2}(x_{2},y_{2})$ corresponding (in the sharp separation picture, Fig.\ref{fig.04}) to these intersections. It reads
\EQ
\Gamma_{ab}(x_{1},y_{1};x_{2},y_{2}\vert u_{1},u_{2}) = \left\{
\begin{array}{ll}
\langle\sigma_{1}\rangle_{a}\langle\sigma_{2}\rangle_{a} \,, \hspace{12mm} \text{if} \quad \text{min}(u_{1},u_{2}) > \text{max}(x_{1},x_{2}) , \\
\\
\langle\sigma_{1}\rangle_{a}\langle\sigma_{2}\rangle_{b} \,, \hspace{12mm} \text{if} \quad u_{1} > x_{1} \wedge u_{2} < x_{2} , \\
\\
\langle\sigma_{1}\rangle_{b}\langle\sigma_{2}\rangle_{a} \,, \hspace{12mm} \text{if} \quad u_{1} < x_{1} \wedge u_{2} > x_{2} , \\
\\
\langle\sigma_{1}\rangle_{b}\langle\sigma_{2}\rangle_{b} \,, \hspace{12mm} \text{if} \quad \text{max}(u_{1},u_{2}) < \text{min}(x_{1},x_{2}),\\
\end{array} 
\right .
\label{Gamma}
\EN
and leads to
\begin{eqnarray} \nonumber
\label{*4.04}
\langle\sigma_{1}(x_{1},y_{1})\sigma_{2}(x_{2},y_{2})\rangle_{ab}^\textrm{sharp}  & = & \langle\sigma_{1}\rangle_{a}\langle\sigma_{2}\rangle_{a} \int_{x_{1}}^{+\infty}\rd u_{1}\int_{x_{2}}^{+\infty}\rd u_{2} \, P_{2} + \langle\sigma_{1}\rangle_{b}\langle\sigma_{2}\rangle_{b} \int_{-\infty}^{x_{1}}\rd u_{1}\int_{-\infty}^{x_{2}}\rd u_{2} \, P_{2} \\ \nonumber
& + & \langle\sigma_{1}\rangle_{a}\langle\sigma_{2}\rangle_{b} \int_{x_{1}}^{\infty}\rd u_{1}\int_{-\infty}^{x_{2}}\rd u_{2} \, P_{2} + \langle\sigma_{1}\rangle_{b}\langle\sigma_{2}\rangle_{a} \int_{-\infty}^{x_{1}}\rd u_{1}\int_{x_{2}}^{+\infty}\rd u_{2} \, P_{2}, \\
\end{eqnarray}
and then to
\begin{equation}
\label{*4.05}
\partial_{x_{1}}\partial_{x_{2}}\langle\sigma_{1}(x_{1},y_{1})\sigma_{2}(x_{2},y_{2})\rangle_{ab}^\textrm{sharp} = \Delta\langle\sigma_{1}\rangle\Delta\langle\sigma_{2}\rangle P_{2}(x_{1},y_{1};x_{2},y_{2}) .
\end{equation}
If, on the other hand, we apply $\partial_{x_{1}}\partial_{x_{2}}$ to (\ref{*3.17}) and equate the result to (\ref{*4.05}) we obtain the expression
\begin{equation}
\label{*4.07}
P_{2}(x_{1},y_{1};x_{2},y_{2}) = \frac{\text{e}^{-\frac{\eta_{1}^{2}}{2(1-\epsilon_{1})}-\frac{\eta_{2}^{2}}{2(1+\epsilon_{2})}-\frac{(\eta_{1}-\eta_{1})^{2}}{2(\epsilon_{1}-\epsilon_{2})}}}{\pi\lambda^{2}\sqrt{2(1-\epsilon_{1})(1+\epsilon_{2})(\epsilon_{1}-\epsilon_{2})}}\,,
\end{equation}
which satisfies the property
\begin{equation}
\label{*4.13}
\int_{\mathbb{R}}\rd u_{2} \, P_{2}(u_{1},y_{1};u_{2},y_{2}) = P_{1}(u_{1},y_{1})
\end{equation}
required for the joint passage probability density (we recall that $P_1$ is given by (\ref{*1.12})). We illustrate in appendix~\ref{App_B} how (\ref{*4.07}) arises in the Brownian bridge picture; it can also be recognized as a bivariate normal distribution\footnote{See (\ref{*A.14}) for the standardized version.} \cite{WR} of the random variables $x_{1},x_{2}$ with covariance matrix
\begin{equation}
\label{*4.08}
\textrm{cov}[x_{1},x_{2}] = \overline{\left(x_{1}-\overline{x}_{1}\right)\left(x_{2}-\overline{x}_{2}\right)} =\frac{1}{2} \left(
\begin{array}{cc}
\kappa_{1}^{2} & \rho \kappa_{1}\kappa_{2} \\
\rho \kappa_{1}\kappa_{2} & \kappa_{2}^{2}
\end{array}\right) ,
\end{equation}
with $\overline{x}_{j}  = 0$ for our case, $\kappa_{j}=\sqrt{1-\epsilon_{j}^{2}}$, and correlation coefficient $\rho$ such that $\rho^{2}=\frac{1-\epsilon_{1}}{1+\epsilon_{1}}\frac{1+\epsilon_{2}}{1-\epsilon_{2}}$. Perfect correlation corresponds to $\rho=1$ and absence of correlation to $\rho=0$; notice however that this limiting cases are never realized within the limits of validity of our field theoretical derivation specified at the beginning of section~\ref{*sec.03}. The probability density can also be written as 
\begin{equation}
\label{*4.09}
P_{2}(x_{1},y_{1};x_{2},y_{2}) = \frac{1}{\pi \kappa_{1}\kappa_{2} \lambda^{2} \sqrt{1-\rho^{2}}}\exp{\biggl[-\frac{\chi_{1}^{2}+\chi_{2}^{2}-2\rho\chi_{1}\chi_{2}}{1-\rho^{2}}\biggr]}\,.
\end{equation}

We notice that an approach based on equations of the type (\ref{*4.01}) and (\ref{Gamma}) was adopted in \cite{Weeks,BW} to obtain an expression for the order parameter two-point function, adopting a Gaussian passage probability density for the interface. The logic of this section is quite different. We have determined the two-point function in two dimensions directly from field theory, and showed that the result is consistent with (\ref{*4.01}) and (\ref{Gamma}) and determines (\ref{*4.07}). 
It is also important to stress that (\ref{*4.01}) accounts only for the leading term of the two-point function in the limits we specified in section~\ref{*sec.03}. Field theory yields also the subleading terms, associated to the internal structure of the interface and to boundary effects. The field theoretical derivation of the first subleading term and its interpretation in terms of interface structure is given in the appendices \ref{app_correlator} and \ref{app_correlator_prob}, respectively; subsequent terms are analyzed in the next section. 

Remaining at leading order and calling {\it height} the deviation $h(y)$ at ordinate $y$ of the position of the interface from the average value $\overline{x}=0$, we obtain the height-height correlation function 
\begin{equation}
\label{hh}
\overline{h(y)h(-y)}= \int_{\mathbb{R}^{2}}\textrm{d}x_{1}\textrm{d}x_{2} \, x_{1}x_{2}\, P_{2}(x_{1},y;x_{2},-y)  = \frac{R}{4m}\,(1-\epsilon)^2\,,
\EN
with $\epsilon=2y/R\ll 1$. In three dimensions the height variable has support on the plane corresponding to minimal interfacial area and is effectively identified with a field which, if massless, has long range correlations. In our two-dimensional case the height has support on a line and cannot properly be treated as a field, so that we can only observe the algebraic  form of the result (\ref{hh}). This can be compared with the form $\overline{h(y)h(-y)}/\overline{h(0)h(0)}=\textrm{e}^{-2y/L_c}$ obtained in \cite{BW} for the two-dimensional case with $R=\infty$ and in presence of an external field $g\propto 1/L_c$. The two forms are formally consistent at leading order if one takes $L_c$ proportional to $R$, and both lengths much larger than the separation $2y$.

\section{Interface structure factor}
\subsection{Connected correlator}
\label{*sec.06}
We begin this section by writing down the connected two-point correlator of the order parameter field. This is obtained from $\langle\sigma_{1}(x_{1},y_{1})\sigma_{2}(x_{2},y_{2})\rangle_{ab}$ through the subtractions ensuring  a vanishing limit when $x_1$ and/or $x_2$ go to infinity. It then reads
\begin{equation}
\label{*6.01}
\langle\sigma_{1}(x_{1},y_{1})\sigma_{2}(x_{2},y_{2})\rangle_{ab}^\textrm{conn} = \Bigl\langle \Bigl[ \sigma_{1}(x_{1},y_{1}) - \mathscr{S}_{ab}(x_{1}) \Bigr] \Bigl[ \sigma_{2}(x_{2},y_{2}) - \mathscr{S}_{ab}(x_{2}) \Bigr] \Bigr\rangle_{ab} - \widetilde{\mathscr{B}}_{ab}(x_{1},y_1;x_{2},y_2) ,
\end{equation}
where
\begin{equation}
\label{*6.02}
\mathscr{S}_{ab}(x) = \langle\sigma_{j}\rangle_{a} \theta(-x) + \langle\sigma_{j}\rangle_{b} \theta(x) ,
\end{equation}
and
\begin{equation}
\label{*6.03}
\widetilde{\mathscr{B}}_{ab}(x_{1},y_1;x_{2},y_2) = \mathscr{B}_{ab}(x_{1},y_1;x_{2},y_2) - \mathscr{S}_{ab}(x_{1})\mathscr{S}_{ab}(x_{2}) ,
\end{equation}
with
\begin{equation}
\label{*6.04}
\mathscr{B}_{ab}(x_{1},y_1;x_{2},y_2) = \langle\sigma_{1}\sigma_{2}\rangle_{a} \theta(-x_{1})\theta(-x_{2}) + \langle\sigma_{1}\sigma_{2}\rangle_{b} \theta(x_{1})\theta(x_{2}) + \mathscr{S}_{ab}(x_{1})\mathscr{S}_{ab}(x_{2})\theta(-x_{1}x_{2})\,;
\end{equation}
$\theta(x)$ is the Heaviside step function. We can also write
\begin{eqnarray} \nonumber
\label{*6.05}
\widetilde{\mathscr{B}}_{ab}(+,+) & = & \langle\sigma_{1}\sigma_{2}\rangle_{b} - \langle\sigma_{1}\rangle_{b}\langle\sigma_{2}\rangle_{b} \equiv \mathscr{G}_{b}(x_{1},y_1;x_{2},y_2) , \\ \nonumber
\widetilde{\mathscr{B}}_{ab}(-,-) & = & \langle\sigma_{1}\sigma_{2}\rangle_{a} - \langle\sigma_{1}\rangle_{a}\langle\sigma_{2}\rangle_{a} \equiv \mathscr{G}_{a}(x_{1},y_1;x_{2},y_2) , \\
\widetilde{\mathscr{B}}_{ab}(\pm,\mp) & = & 0 ,\nonumber
\end{eqnarray}
where $\pm$ refer to the sign of the $x_i$'s, and $\mathscr{G}_{\mu}$ is the connected bulk correlator for the pure phase $\mu$. The subtraction of $\mathscr{B}_{ab}$ in (\ref{*6.01}) eliminates the bulk term one obtains when $x_1$ and $x_2$ are simultaneously translated to infinity keeping the relative distance fixed. Within our large distance expansion over kink intermediate states the first contributions of this type to $\langle\sigma_{1}\sigma_{2}\rangle_{ab}$ have the pictorial representation 
\begin{equation}
\label{*6.13}
\newfig{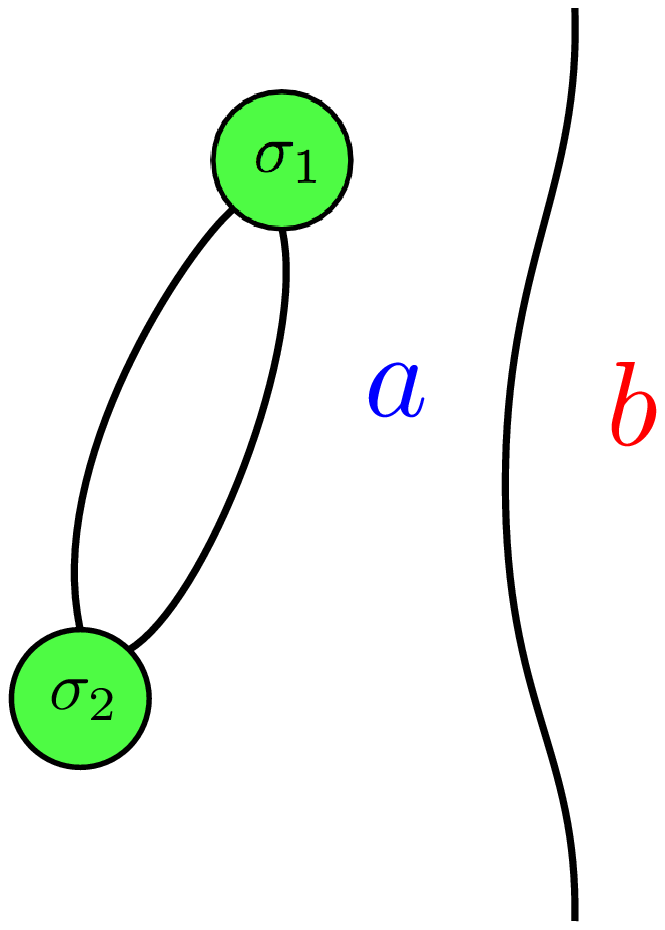} \hspace{15mm} , \hspace{35mm} \newfig{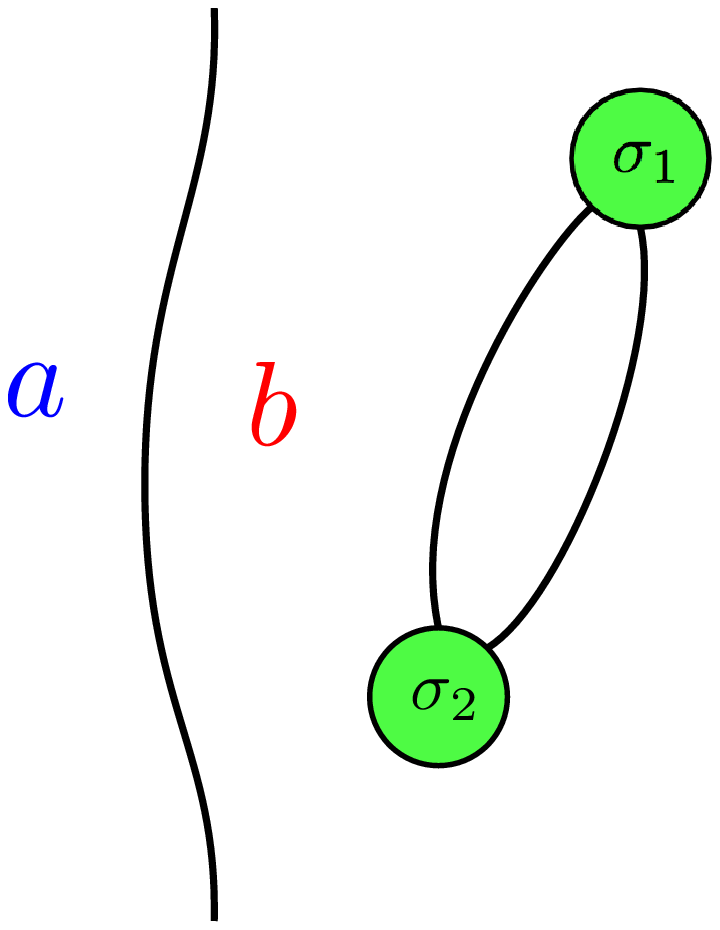} \hspace{15mm},
\end{equation}
and correspond to a three-kink intermediate state. Hence they are definitely subleading with respect to the single kink term we are analyzing. This is why we can ignore the term $\widetilde{\mathscr{B}}_{ab}$ in the following. Concerning the remaining part of (\ref{*6.01}), i.e.
\begin{equation}
\label{*6.06}
\langle\sigma_{1}(x_{1},y_{1})\sigma_{2}(x_{2},y_{2})\rangle_{ab} - \mathscr{S}_{ab}(x_{1}) \langle\sigma_{2}(x_{2},y_{2})\rangle_{ab} - \mathscr{S}_{ab}(x_{2}) \langle\sigma_{1}(x_{1},y_{1})\rangle_{ab} + \mathscr{S}_{ab}(x_{1}) \mathscr{S}_{ab}(x_{2})\,,
\end{equation}
we can use (\ref{*3.16}) to see that at the one-kink level it reduces to (\ref{*3.09}) plus terms which are odd in at least one of the variables $x_1$ and $x_2$. Since these odd terms give a vanishing contribution upon integration over $x_i$ in (\ref{sf}), we arrive at the conclusion 
\begin{equation}
\label{*6.14}
\hat{S}(q) = \frac{1}{2({\langle\sigma\rangle_a-\langle\sigma\rangle_b})^2}\int_{-R/2}^{R/2}\textrm{d}y \, \text{e}^{-iqy} \lim_{L\rightarrow\infty} \int_{-L}^{L}\textrm{d}x_{1}\int_{-L}^{L}\textrm{d}x_{2} \, \langle\sigma(x_{1},y)\sigma(x_{2},-y)\rangle_{ab}^\textrm{CP}\,;
\end{equation}
it follows from (\ref{*3.18}) and (\ref{*3.19}) that $\langle\sigma\sigma\rangle_{ab}^\textrm{CP}$ tends to opposite (and generically non-zero) values when $x_i$ goes to plus or minus infinity, and the integration over the symmetric interval $(-L,L)$ yields a convergent result for $L\to\infty$.

\subsection{Large $R$ expansion}
\label{*sec.07}
The expressions (\ref{*3.09}), (\ref{*3.21}) determine only the leading term of $\langle\sigma\sigma\rangle_{ab}^\textrm{CP}$ for large $R$. In the following we will also consider the corrections generated by subsequent terms in the small rapidity expansions of the connected part of the matrix element (\ref{*1.06}), and of the boundary amplitude $f_{ab}$ entering (\ref{*1.01}). We write these expansions as
\EQ
{\cal F}_{ab}^\sigma(\theta)=\Delta\langle\sigma\rangle\sum_{k=-1}^\infty c_k\theta^k\,,
\label{ff_expansion}
\EN
which extends (\ref{*1.06}), and 
\EQ
f_{ab}(\theta)=1+\sum_{k=1}^{\infty}\alpha_{2k}\theta^{2k}\,;
\label{boundary_expansion}
\EN
when expanding (\ref{boundary_expansion}) over even powers we restrict, for the sake of simplicity, to cases in which the phases $a$ and $b$ play a symmetric role, as for Ising and Potts universality classes. 
It is a consequence of (\ref{*3.04}) and (\ref{*3.06}) that (\ref{ff_expansion}) and (\ref{boundary_expansion}) will induce a large $R$ expansion of the correlator with suppression factors of the form $(mR)^{-\frac{\ell}{2}}$, with $\ell\geq 0$, and we write
\begin{equation}
\label{*7.01}
\langle\sigma_{1}(x_{1};y)\sigma_{2}(x_{2};-y)\rangle_{ab}^{\textrm{CP}} = \sum_{\ell=0}^{\infty} \bigl[ \langle\sigma_{1}\sigma_{2}\rangle^{\textrm{CP}} \bigl]_{\ell} (mR)^{-\ell/2}\,;
\end{equation}
also for $\ell>0$ the quantities $\bigl[ \langle\sigma_{1}\sigma_{2}\rangle^{\textrm{CP}} \bigl]_{\ell}$ have a constant limit for $R\to\infty$, and we now turn to the determination of the first few of them. 

Besides (\ref{boundary_expansion}) we also use $|f_{ab}(\theta)|^{2}=1+\sum_{k=1}^{\infty} f_{2k}\theta^{2k}$, and write the partition function (\ref{partition}) beyond leading order for large $R$ as
\begin{equation}
\label{*7.06}
\mathcal{Z}_{ab}\simeq\int_{\mathbb{R}} \frac{\rd\theta}{2\pi} \vert f_{ab}(\theta) \vert^{2} \text{e}^{-mR\cosh\theta}\simeq\mathcal{Z}_{ab}^{(0)} \biggl[ 1+\sum_{k=1}^\infty \zeta_{k}(mR)^{-k}\biggr]\, ,
\end{equation}
where $\mathcal{Z}_{ab}^{(0)}=\frac{\text{e}^{-mR}}{\sqrt{2\pi mR}}$, and $\zeta_{k} = (2k-1)!! f_{2k}$; in particular $\zeta_{1}=2\alpha_{2}$.

We now turn to the expansion of the numerator of (\ref{*3.01}). Concerning the contribution to $\langle\sigma_{1}\sigma_{2}\rangle^{\textrm{CP}}$  we have to expand the quantity
\begin{equation}
\label{*7.07}
Q(\theta_{1},\theta_{2},\theta_{3}) \equiv \overline{f_{ab}}(\theta_{1})f_{ab}(\theta_{2}) \mathcal{F}_{ab}^{\sigma_{1}}(\theta_{1}\vert\theta_{3}) \mathcal{F}_{ab}^{\sigma_{2}}(\theta_{3}\vert\theta_{2}) ,
\end{equation}
for small $\theta_{j}$, $j=1,2,3$, and evaluate the first terms of such an expansion. Each term is a monomial in the rapidities of the form $\theta_{1}^{p}\theta_{2}^{q}$, with $p$ and $q$ even and non-negative, multiplied by $\theta_{13}^{r}\theta_{32}^{s}$ (we recall (\ref{theta12})). It follows that such a term is a homogeneous function in the rapidities with degree $\Delta=p+q+r+s$, with $\Delta+2\in\mathbb{N}$. The leading term is characterized by the minimum value of the homogeneity exponent, $\Delta=-2$, while the first subleading correction to the two-point function comes from $\Delta=-1$. Let us use the following shorthand notation 
\EQ
Q(\theta_{1},\theta_{2},\theta_{3}) = \sum_{\Delta=-2}^{+\infty} Q_{\Delta}\,,
\label{QDelta}
\EN
in which $Q_{\Delta}$ denotes all the terms with the same homogeneity exponent.
It is easy to see that a term characterized by a certain $\Delta$ will produce a factor $(mR)^{-\frac{3+\Delta}{2}}$ in the numerator of (\ref{*3.01}). The leading term $\Delta=-2$ produces a factor $(mR)^{-\frac{1}{2}}$ which is cancelled by $\mathcal{Z}_{ab}^{(0)}$ in the denominator. Therefore we can write
\begin{equation}
\label{*7.09}
\langle\sigma_{1}(x_{1};y)\sigma_{2}(x_{2};-y)\rangle_{ab}^{\text{CP}}\simeq \frac{\frac{1}{\mathcal{Z}_{ab}^{(0)}} \sum_{\Delta=-2}^{+\infty} \int_{\mathbb{R}^{3}}\frac{\textrm{d}\theta_{1}\textrm{d}\theta_{2}\textrm{d}\theta_{3}}{(2\pi)^{3}} Q_{\Delta} \widetilde{\mathcal{E}}(\theta_{1},\theta_{2},\theta_{3})}{ 1 + \sum_{k=1}^{\infty}\frac{\zeta_{k}}{(mR)^{k}}}\, ,
\end{equation}
with $\widetilde{\mathcal{E}}$ given by (\ref{*3.06}). Now we rescale the rapidities as $\theta_{j}\rightarrow \sqrt{\frac{2}{mR}}\theta_{j}$, define 
\begin{equation}
\label{*7.10}
Y_{\epsilon}(\theta_{1},\theta_{2},\theta_{3}) \text{e}^{i\eta_{1}\theta_{13}+i\eta_{2}\theta_{32}} \equiv \textrm{e}^{mR}\widetilde{\mathcal{E}}\left(\sqrt{\frac{2}{mR}}\theta_{1},\sqrt{\frac{2}{mR}}\theta_{2},\sqrt{\frac{2}{mR}}\theta_{3}\right) ,
\end{equation}
\EQ
Y_{\epsilon}(\theta_{1},\theta_{2},\theta_{3}) \equiv \text{e}^{-\frac{1-\epsilon}{2}\left(\theta_{1}^{2}+\theta_{2}^{2}\right)-\epsilon \theta_{3}^{2}}\,,
\EN
and introduce the shorthand notation
\begin{equation}
\label{*7.11}
\{ \Phi \} \equiv \int_{\mathbb{R}^{3}}\textrm{d}\theta_{1}\textrm{d}\theta_{2}\textrm{d}\theta_{3} \, \Phi(\theta_{1},\theta_{2},\theta_{3}) \, Y_{\epsilon}(\theta_{1},\theta_{2},\theta_{3}) \text{e}^{i\eta_{1}\theta_{13}+i\eta_{2}\theta_{32}}\,;
\end{equation}
$\bigl\{ \Phi \bigr\}$ is a scaling function of the dimensionless variables $\eta_{1},\eta_{2},\epsilon$ (recall (\ref{etaeps})). These manipulations allow us to write
\begin{equation}
\label{*7.12}
\langle\sigma_{1}(x_{1};y)\sigma_{2}(x_{2};-y)\rangle_{ab}^{\text{CP}}\simeq \frac{\sum_{\Delta=-2}^{+\infty}\bigl\{\frac{Q_{\Delta}}{4\pi^{5/2}|a_{0}|^{2}}\bigr\} \left(\frac{2}{mR}\right)^{\frac{\Delta+2}{2}}}{1 + \sum_{k=1}^{\infty}\frac{\zeta_{k}}{(mR)^{k}}}\,,
\end{equation}
and then the expansion in powers of $R^{-1/2}$
\begin{eqnarray} \nonumber
\langle\sigma_{1}(x_{1};y)\sigma_{2}(x_{2};-y)\rangle_{ab}^{\text{CP}} &\simeq & \biggl\{\frac{Q_{-2}}{4\pi^{5/2}}\biggr\} + \biggl\{\frac{Q_{-1}}{4\pi^{5/2}}\biggr\}\sqrt{\frac{2}{mR}} + \\ \nonumber
& + & \Biggl[ \biggl\{\frac{Q_{0}}{2\pi^{5/2}}\biggr\} - \biggl\{\frac{\zeta_{1}Q_{-2}}{4\pi^{5/2}}\biggr\} \Biggr] \frac{1}{mR} + \bigg[ \dots \biggr] \frac{1}{(mR)^{3/2}} + \\ \nonumber
& + &\Biggl[ \biggl\{\frac{Q_{2}}{\pi^{5/2}}\biggr\} - \biggl\{\frac{\zeta_{1}Q_{0}}{2\pi^{5/2}}\biggr\} +  \biggl\{\frac{\zeta_{1}^{2}-\zeta_{2}}{4\pi^{5/2}}Q_{-2}\biggr\} \Biggr] \frac{1}{(mR)^{2}} + \\
& + & \mathcal{O}\left( R^{-\frac{5}{2}} \right)\,,
\label{*7.13}
\end{eqnarray}
which corresponds to (\ref{*7.01}); we did not write explicitly the factor multiplying $R^{-3/2}$ because, for the parity arguments that we are going to discuss, it does not contribute to the structure factor.

The calculation of the interface structure factor (\ref{*6.14}) requires the integration of (\ref{*7.13}) over $x_{1},x_{2}$. It will be convenient to introduce a compact notation for the spatial integral of a scaling function. Thus, given a function $\Phi$ of the rapidities we construct the associated scaling function thanks to (\ref{*7.11}), and the spatial integral as
\begin{equation}
\label{*7.14}
\llbracket \Phi \rrbracket \equiv \lim_{\Lambda\rightarrow\infty}\int_{-\Lambda}^{\Lambda}\textrm{d}\eta_{1}\int_{-\Lambda}^{\Lambda}\textrm{d}\eta_{2} \, \Bigl\{ \Phi(\theta_{1},\theta_{2},\theta_{3}) \Bigr\} \Big\vert_{\epsilon\rightarrow|\epsilon|}\,.
\end{equation}
Since the time ordering of the fields adopted so far implied $y>0$, we perform the replacement $\epsilon\rightarrow|\epsilon|$ in order to have the result which holds also for $y<0$; the factor $1/2$ in (\ref{sf}) avoids double counting when integrating over positive and negative values of $y$. We then further define
\begin{equation}
\label{*7.15}
\widehat{\llbracket \Phi \rrbracket} \equiv \int_{-1}^{1}\textrm{d}\epsilon \, \llbracket \Phi \rrbracket \, \text{e}^{-iQ\epsilon}\,,
\end{equation}
\EQ
Q=\frac{qR}{2}\,.
\label{Q}
\EN

The following Lemmas prove to be useful in view of the calculation of $\hat{S}(q)$. 

\begin{itemize}
\item \noindent\textbf{Lemma 1}\\
The integration of $\bigl\{Q_{\Delta}\bigr\}$ over $(\eta_{1},\eta_{2}) \in (-\Lambda,\Lambda) \times (-\Lambda,\Lambda)$ vanishes if $\Delta$ is an odd integer.

\noindent\emph{Proof:}\\
It follows from the definition (\ref{*7.11}) that $(\eta_1,\eta_2)\to(-\eta_1,-\eta_2)$ is related to $(\theta_1,\theta_2,\theta_3)\to(-\theta_1,-\theta_2,-\theta_3)$, and leads to $\bigl\{Q_{\Delta}\bigr\}\to(-1)^{\Delta}\bigl\{Q_{\Delta}\bigr\}$. Then the integration over $\eta_i$ vanishes for $\Delta$ odd.
\begin{flushright}
\vspace{-10mm}
$\boxed{}$
\end{flushright}
\end{itemize}

\begin{itemize}
\item \noindent\textbf{Lemma 2}\\
The functions $\theta_{1}^{p}\theta_{2}^{q}\theta_{13}^{r}\theta_{32}^{s}$ with $\max_{\Omega}(r,s)\geqslant1$ and $\Omega=\{(r,s) \vert r+1,s+1 \in \mathbb{N}^{2} \}$ give rise to $\llbracket \theta_{1}^{p}\theta_{2}^{q}\theta_{13}^{r}\theta_{32}^{s} \rrbracket = 0$.

\noindent\emph{Proof:}\\
Let us consider the function $f(\theta_{1},\theta_{2})\theta_{13}^{r}\theta_{32}^{s}$ with $f(\theta_{1},\theta_{2})=\theta_{1}^{p}\theta_{2}^{q}$ in the following cases:
\begin{enumerate}
\item[(a):] $\min(r,s)\geqslant1$;
\item[(b):] $(r,s)=(0,s)$ with $s\geqslant1$ or $(r,s)=(r,0)$ with $r\geqslant1$;
\item[(c):] $(r,s)=(-1,s)$ with $s\geqslant1$ or $(r,s)=(r,-1)$ with $r\geqslant1$.
\end{enumerate}
For simplicity we can examine case (a) corresponding to $r,s\geqslant1$. We have
\begin{equation}
\label{*7.17}
\int_{\mathbb{R}^{2}}\textrm{d}\eta_{1}\textrm{d}\eta_{2} \, \Bigl\{ f(\theta_{1},\theta_{2}) \theta_{13}^{r}\theta_{32}^{s} \Bigr\} = (2\pi)^{2} \Bigl\{ f(\theta_{1},\theta_{2})\theta_{13}^{r}\theta_{32}^{s}\delta(\theta_{13})\delta(\theta_{32}) \Bigr\} = 0 ,
\end{equation}
due to the Dirac delta functions. We reach the same conclusions even when we have a single delta function, corresponding to cases (b) and (c).
\begin{flushright}
\vspace{-10mm}
$\boxed{}$
\end{flushright}
\end{itemize}

\begin{itemize}
\item \noindent\textbf{Corollary 1}\\
The functions $Y_{p,q}^{(0,0)}=\theta_{1}^{p}\theta_{2}^{q}$ and $Y_{p,q}^{(-1,0)}=\theta_{1}^{p}\theta_{2}^{q}\theta_{13}^{-1}$, $Y_{p,q}^{(0,-1)}=\theta_{1}^{p}\theta_{2}^{q}\theta_{32}^{-1}$ with $(p,q)\in\mathbb{N}^{2}$ give rise to $\llbracket Y_{p,q}^{(0,0)} \rrbracket = 0$ if $p+q$ is odd and $\llbracket Y_{p,q}^{(-1,0)} \rrbracket = \llbracket Y_{p,q}^{(0,-1)} \rrbracket  = 0$ if $p+q$ is even.

\noindent\emph{Proof:}\\
If $p+q$ is odd then $\{Y_{p,q}^{(0,0)}\}$ is odd, while if $p+q$ is even $\{Y_{p,q}^{(-1,0)}\}$ and $\{Y_{p,q}^{(0,-1)}\}$ are odd, therefore by virtue of Lemma 1 their integral over the spatial coordinates vanishes.
\begin{flushright}
\vspace{-10mm}
$\boxed{}$
\end{flushright}
\end{itemize}

\begin{itemize}
\item \noindent\textbf{Corollary 2}\\
For our case (\ref{boundary_expansion}) the only non-vanishing contributions to $\hat{S}(q)$ come from terms with zero or two poles, namely of the form $f_{ab}(\theta_{1})f_{ab}(\theta_{2})$ and $\frac{f_{ab}(\theta_{1})f_{ab}(\theta_{2})}{\theta_{13}\theta_{32}}$.

\noindent\emph{Proof:}\\
$p+q$ is an even integer, therefore a term with $\theta_{13}^{-1}$ as the only pole gives rise to $\Delta=p+q-1$ which is odd. Hence, thanks to Lemma 1 its contribution to $\hat{S}(q)$ vanishes. The pole-free and the double-pole terms $f_{ab}(\theta_{1})f_{ab}(\theta_{2})$ and $\frac{f_{ab}(\theta_{1})f_{ab}(\theta_{2})}{\theta_{13}\theta_{32}}$ have even $\Delta$ and survive Lemma~1.
\begin{flushright}
\vspace{-10mm}
$\boxed{}$
\end{flushright}
\end{itemize}

In summary, with reference to (\ref{ff_expansion}), the non-vanishing contributions to $\hat{S}(q)$ will be those proportional to $c_{-1}^{2}$ and $c_{0}^{2}$. Recalling (\ref{*1.06}) we know that $c_{-1}^2=-1$; the vanishing of the contributions containing $c_j$ with $j>0$ is not obvious a priori.

We write the interfacial structure factor (\ref{*6.14}) as 
\begin{equation}
\label{*7.02}
\hat{S}(q)\simeq \sum_{\ell=0}^{\infty} \hat{S}_{\ell}(q)\,,
\end{equation}
where $\hat{S}_{\ell}(q)$ is the contribution of the term proportional to $R^{-\ell/2}$ in (\ref{*7.13}). The terms $\hat{S}_{\ell}(q)$ with $\ell$ odd vanish by virtue of Lemma 1, hence we focus on those with $\ell$ even.
Recalling also (\ref{*7.14}) and (\ref{*7.15}), the first term in (\ref{*7.02}) is
\EQ
\hat{S}_{0}(q)=-\frac{R^{2}}{32\pi^{5/2}m} \widehat{\llbracket \tau_{-2,1} \rrbracket}=\frac{1}{mq^{2}} \left( 1 - \frac{\sin Q}{Q} \right)
\label{*8.03}
\EN
where $\tau_{-2,1}\equiv 1/(\theta_{13}\theta_{32})$ is proportional to $Q_{-2}$. The functions $\tau_{\Delta,j}$, as well as their integrated form $\widehat{\llbracket \tau_{\Delta,j} \rrbracket}$, are listed in appendix~C. Using Lemma~2 to get rid of some contributions coming from $Q_0$ and $Q_2$, we can further write
\begin{eqnarray}
\nonumber
\hat{S}_{2}(q) & = & \frac{\lambda^{2}R/2}{2mR} \frac{1}{2\pi^{5/2}} \biggl[ c_{0}^{2} \widehat{\llbracket \tau_{0,1} \rrbracket} - \alpha_{2}\widehat{\llbracket \tau_{0,2} \rrbracket} + \frac{\zeta_{1}}{2} \widehat{\llbracket \tau_{-2,1} \rrbracket} \bigg]\\
& = & \frac{c_{0}^{2}\sin Q}{m^{2}q} + 4\alpha_{2}\frac{\cos Q}{m^{2}q^{2}R} + \mathcal{O}\left( R^{-2} \right),
\label{*8.05}
\end{eqnarray}
\begin{eqnarray}
\nonumber
\hat{S}_{4}(q) & = & \frac{\lambda^{2}R/2}{2m^{2}R^{2}} \frac{1}{\pi^{5/2}} \biggl[ \alpha_{2}c_{0}^{2}\widehat{\llbracket \tau_{2,1} \rrbracket} -\alpha_{4}\widehat{\llbracket \tau_{2,2} \rrbracket} - \alpha_{2}^{2}\widehat{\llbracket \tau_{2,3} \rrbracket} \biggr] + \\ \nonumber
& + & \frac{\lambda^{2}R/2}{2m^{2}R^{2}} \frac{-\zeta_{1}}{2\pi^{5/2}} \biggl[ c_{0}^{2} \widehat{\llbracket \tau_{0,1} \rrbracket} - \alpha_{2} \widehat{\llbracket \tau_{0,2} \rrbracket} \biggr] + \frac{\lambda^{2}R/2}{2m^{2}R^{2}} \frac{\zeta_{1}^{2}-\zeta_{2}}{4\pi^{5/2}} \biggl[ - \widehat{\llbracket \tau_{-2,1} \rrbracket} \biggr]\nonumber\\
& = & 4\alpha_{2}^{2} \frac{\sin Q}{m^{3}qR} + \mathcal{O}\left( R^{-2} \right)\,;
\label{*8.07} 
\end{eqnarray}
It can be checked that the terms $\hat{S}_{2k}(q)$ with $k>2$ do not contribute to order $1/R$. Putting together these results and those of appendix~C we obtain (\ref{*8.16}), where we have to consider $q$ much smaller than $m$ and larger than $q_0\propto 1/R$.

\section{Conclusion}
\label{*sec.09}
In this paper we considered two-dimensional systems at phase coexistence near a second order phase transition point and determined the form of the long range order parameter correlations. We were able to do this in an exact way through the extension to two-point functions of the field theoretical formalism developed in \cite{DV,DS2}. More precisely, we considered an infinitely long strip of width $R$ much larger than the bulk correlation length, and with boundary conditions which induce the separation of two phases and an interface running from one edge to the other. We then showed that, as long as $R$ is finite, the order parameter has long range correlations of the specific form (\ref{leading}) in the $y$-direction parallel to the interface. For $R=\infty$ the fluctuations of the interface become infinitely wide and leave only exponentially decaying bulk correlations averaged over the two phases. 

Technically, a key role is played by the fact that for phase separation in two dimensions the excitations of the underlying field theory have a topological nature (they are kinks), and are non-local with respect to the order parameter field, a fact which reflects into the singularity (\ref{*1.06}) in the matrix element of the order parameter. Singularities of a similar nature exist and play an important role also in higher dimensions \cite{vortex}, but in that case they are not related to phase separation. 

We also determined in field theory subleading corrections to the large $R$ expansion of the two-point function. We showed that the leading term amounts to the presence of an interface behaving as a simple curve which sharply separates two pure phases and fluctuates according to a Gaussian passage probability density. Subleading corrections then correspond to endowing the interface with an internal structure.

Our results for the order parameter two-point function allowed us also a direct investigation of the structure factor of the interface. This quantity depends on a single variable and is largely considered in the framework of effective descriptions aiming at a compact characterization of the interfacial properties. We showed how the term proportional to $1/q^2$, which in momentum space is the signature of long range correlations, emerges from the expression of the two-point function in real space. The specific form of the latter also characteristically manifests into $R$-dependent corrections which depend on bulk and boundary data and localize towards $q=0$ as $R\to\infty$.

\section*{Acknowledgements}
AS thanks for their kind hospitality the organizers of the workshop ``Soft matter at interfaces 2016'', Ringberg Castle (Tegernsee), during which parts of this work were done.

\appendix
\section{Integrals}
\label{App_A}
In this appendix we collect the integrals needed in the main text and some other useful mathematical result. Owen's $T$ function is defined through the integral
\begin{equation}
\label{*A.01}
T(h,a) = \frac{1}{2\pi}\int_{0}^{a}\rd x \, \frac{\text{e}^{-h^{2}\frac{1+x^{2}}{2}}}{1+x^{2}} ,
\end{equation}
and satisfies $T(h,a)=T(-h,a)=-T(h,-a)$ and $T(h,0)=T(\pm\infty,a)=0$. For special values of the arguments Owen's $T$ function reduces to elementary functions
\begin{eqnarray} \nonumber
\label{*A.02}
T(0,a) & = & \frac{\tan^{-1}(a)}{2\pi} , \\ \nonumber
T(\sqrt{2}h,1) & = & \frac{1-\text{erf}^{2}(h)}{8} = \frac{1}{8}\text{erfc}\left(-h\right)\text{erfc}\left(h\right), \\
T(\sqrt{2}h,\pm\infty) & = & \pm\frac{1}{4}\text{erfc}\left(|h|\right) ,
\end{eqnarray}
where $\text{erf}(x)=\frac{2}{\sqrt{\pi}}\int_{0}^{x} \rd u \, \text{e}^{-u^{2}}$ is the error function and $\text{erfc}(x)=1-\text{erf}(x)$ is the complementary error function. The above expressions are useful in the study of the asymptotic properties of (\ref{*3.21}). The function $T$ obeys also the functional equation
\begin{equation}
\label{*A.03}
T(h,a) + T(ah,a^{-1}) = \frac{1}{2}\Psi(h) + \frac{1}{2}\Psi(ah) - \Psi(h)\Psi(ah) - \frac{1}{2}\theta(-a) ,
\end{equation}
where $\Psi(x)\equiv (1/2)\textrm{erfc}(-x/\sqrt{2})$ and $\theta(x)$ is Heaviside step function. Owen's $T$ function allows us to write the relation
\begin{equation}
\label{*A.04}
\frac{1}{\sqrt{\pi}}\int_{-\infty}^{x} \rd u \, \text{e}^{-u^{2}}\text{erf}(q u) = -2T(\sqrt{2}x,q) ,
\end{equation}
which will be used during the subsequent manipulations. Another result needed in the main body of the paper is
\begin{equation}
\int\rd\eta \, \text{e}^{-\frac{\eta^{2}}{2a}+i\eta\theta} = \sqrt{\frac{\pi a}{2}} \, \text{e}^{-\frac{a\theta^{2}}{2}} \, \text{erf}\left(\frac{\eta-ia\theta}{\sqrt{2a}}\right) + C\,,
\end{equation}
with $C$ an arbitrary constant.

The remaining part of this appendix is devoted to prove the integrals listed below:
\begin{eqnarray}
\label{*A.05}
I(a,b) & = & \frac{1}{\sqrt{\pi}} \int_{\mathbb{R}} \rd u \, \text{e}^{-u^{2}} \text{erf}(au+b) = \text{erf}\left(\frac{b}{\sqrt{1+a^{2}}}\right) , \\ \nonumber
J_{\pm}(a,b) & = & \frac{1}{\sqrt{\pi}} \int_{\mathbb{R}} \rd u \, \text{e}^{-u^{2}} \text{erf}(au+b)\text{erf}(\pm au + b) = 1 - 8 T\left(\sqrt{\frac{2}{1+a^{2}}}b,\left(1+2a^{2}\right)^{\mp\frac{1}{2}} \right) , \\
\label{*A.06}
&& \\
\label{*A.07}
F(a_{1},b_{1},a_{2}) & = & \frac{1}{\sqrt{\pi}} \int_{\mathbb{R}} \rd u \, \text{e}^{-u^{2}} \text{erf}(a_{1}u+b_{1})\text{erf}(a_{2}u) = 4 T\left(\sqrt{\frac{2}{1+a_{1}^{2}}}b_{1},\frac{a_{1}a_{2}}{\sqrt{1+a_{1}^{2}+a_{2}^{2}}}\right) , \\ \nonumber
\label{*A.08}
G(a_{1},b_{1},a_{2},b_{2}) & = & \frac{1}{\sqrt{\pi}} \int_{\mathbb{R}} \rd u \, \text{e}^{-u^{2}} \text{erf}(a_{1}u+b_{1})\text{erf}(a_{1}u+b_{1}) =  \text{sign}(b_{1}b_{2}) + \\
& - &  4T\left(x_{1},\frac{x_{2}/x_{1}-\rho}{\sqrt{1-\rho^{2}}}\right) - 4T\left(x_{2},\frac{x_{1}/x_{2}-\rho}{\sqrt{1-\rho^{2}}}\right) , \hspace{10mm} \text{for} \,\, b_{1}b_{2}\neq 0,
\end{eqnarray}
with $x_{j}=\sqrt{\frac{2}{1+a_{j}^{2}}}b_{j}$ and $\rho =\frac{a_{1}a_{2}}{\sqrt{\left(1+a_{1}^{2}\right)\left(1+a_{2}^{2}\right)}}$. We stress that (\ref{*A.08}) holds for $b_{1}b_{2}\neq0$, while for $b_{1}b_{2}=0$ the function $G$ reduces to the function $F$ \footnote{For completeness we mention that it is possible to define $G$ for arbitrary values of $b_{1},b_{2}$ but the price is that this requires a careful definition of the corresponding Heaviside step function at zero arguments. We prefer to avoid such a complication in favor of (\ref{*A.08}) supplemented by (\ref{*A.07}).}. The function $J_{\pm}$ can be derived from (\ref{*A.08}), and is particularly useful since (\ref{leading}) follows directly from $J_{-}(i\sqrt{(1-\epsilon)/2},\eta/\sqrt{2(1-\epsilon)})$. The integral $I(a,b)$ can be easily performed taking the first derivative with respect to $b$ which produces a Gaussian integral; then integrating over $b$ and using the condition $I(a,0)=0$ we obtain (\ref{*A.05}). Let us consider the function $F$; taking the first derivative with respect to $b_{1}$ and completing the square in the exponential we find an integral analogous to $I$. Thus we can write
\begin{equation}
\label{*A.09}
\partial_{b_{1}}F(a_{1},b,a_{2}) = -\frac{2}{\sqrt{\pi}} \frac{\text{e}^{-\frac{b_{1}^{2}}{1+a_{1}^{2}}}}{\sqrt{1+a_{1}^{2}}}\text{erf}\left(\frac{a_{1}a_{2}b_{1}}{\sqrt{\left(1+a_{1}^{2}\right)\left(1+a_{1}^{2}+a_{2}^{2}\right)}} \right)\,;
\end{equation}
integrating over $b_{1}$ thanks to (\ref{*A.04}) and using $F(a_{1},-\infty,a_{2})=0$ we find
\begin{equation}
\label{*A.10}
F(a_{1},b_{1},a_{2}) = -\frac{2}{\sqrt{\pi}} \int_{-\infty}^{x_{1}/\sqrt{2}}\rd x \, \text{e}^{-x^{2}} \text{erf}\left(\frac{a_{1}a_{2}x}{\sqrt{1+a_{1}^{2}+a_{2}^{2}}}\right) = 4 T\left(\sqrt{\frac{2}{1+a_{1}^{2}}}b_{1},\frac{a_{1}a_{2}}{\sqrt{1+a_{1}^{2}+a_{2}^{2}}}\right) ,
\end{equation}
which proves (\ref{*A.07}). The same strategy can be followed for the integral $G$, where, applying the same techniques of the previous computation, we find
\begin{equation}
\label{*A.11}
\partial_{b_{1},b_{2}}^{2}G(a_{1},b_{1},a_{2},b_{2}) = \frac{4}{\pi\sqrt{1+a_{1}^{2}+a_{2}^{2}}} \text{e}^{-\frac{1+a_{2}^{2}}{1+a_{1}^{2}+a_{2}^{2}}b_{1}^{2} + \frac{2a_{1}a_{2}b_{1}b_{2}}{1+a_{1}^{2}+a_{2}^{2}} - \frac{1+a_{1}^{2}}{1+a_{1}^{2}+a_{2}^{2}}b_{2}^{2}};
\end{equation}
we notice that the right hand side of (\ref{*A.11}) is proportional to a bivariate normal distribution of the random variables $b_{1}$, $b_{2}$. In order to enlighten this connection we introduce the following parametrization
\begin{eqnarray} \nonumber
\label{*A.12}
\sigma_{j}^{2} & = & \frac{1+a_{j}^{2}}{2} , \\
\rho^{2} & = & \frac{a_{1}^{2}a_{2}^{2}}{\left(1+a_{1}^{2}\right)\left(1+a_{2}^{2}\right)} ,
\end{eqnarray}
which allows us to write (\ref{*A.11}) in the form
\begin{equation}
\label{*A.13}
\partial_{b_{1},b_{2}}^{2}G(a_{1},b_{1},a_{2},b_{2}) = \frac{4}{\pi\sqrt{1+a_{1}^{2}+a_{2}^{2}}} \text{e}^{-\frac{1}{2(1-\rho^{2})}\Bigl[\frac{b_{1}^{2}}{\sigma_{1}^{2}}-2\rho\frac{b_{1}b_{2}}{\sigma_{1}\sigma_{2}}+\frac{b_{2}^{2}}{\sigma_{2}^{2}}\Bigr]} .
\end{equation}
We note that $x_{j}=b_{j}/\sigma_{j}$ and therefore it is straightforward to identify in the r.h.s of (\ref{*A.13}) a bivariate normal distribution in the standard form
\begin{equation}
\label{*A.14}
P_{2}(x_{1},x_{2};\rho) = \frac{1}{2\pi\sqrt{1-\rho^{2}}}\text{e}^{-\frac{x_{1}^{2}-2\rho x_{1}x_{2}+x_{2}^{2}}{2(1-\rho^{2})}} ,
\end{equation}
where $\rho=\frac{\langle x_{1} x_{2} \rangle}{\sqrt{\langle x_{1}^{2} \rangle \langle x_{2}^{2} \rangle}}$ is the correlation coefficient. We recall that $\langle x_{j} \rangle = 0$ and $\langle x_{j}^{2} \rangle = 1$ for the standardized distribution (\ref{*A.14}). With the aid of (\ref{*A.14}) we can write (\ref{*A.13}) in the compact form
\begin{equation}
\label{*A.15}
\partial_{x_{1},x_{2}}^{2} G(a_{1},b_{1},a_{2},b_{2}) = 4 P_{2}(x_{1},x_{2};\rho) ,
\end{equation}
therefore the function $G$ can be obtained upon integrating over $x_{1}$ and $x_{2}$ the joint probability $P_{2}$ with the correct asymptotic conditions. It is obvious that this operation corresponds to the cumulative distribution $\Phi$ associated to (\ref{*A.14}); the latter can be written in terms of Owen's function \cite{Owen}
\begin{eqnarray} \nonumber
\Phi(x_{1},x_{2};\rho) & = & \int_{-\infty}^{x_{1}}\rd u_{1}\int_{-\infty}^{x_{2}}\rd u_{2} \, P_{2}(u_{1},u_{2};\rho) \\ \nonumber
& = & \Theta(x_{1},x_{2}) + \frac{\Psi(x_{1})+\Psi(x_{2})}{2} - T\left(x_{1},\frac{x_{2}/x_{1}-\rho}{\sqrt{1-\rho^{2}}}\right) - T\left(x_{2},\frac{x_{1}/x_{2}-\rho}{\sqrt{1-\rho^{2}}}\right) , \\
\end{eqnarray}
where $\Theta(x_{1},x_{2})=\frac{\text{sign}(x_{1}x_{2})-1}{4}$ for $x_{j}\neq0$. It is easy to check that the above reduces to the cumulative distribution for a single random variable if one of the arguments tends to infinity, i.e. $\Phi(x_{1},+\infty;\rho) = \Psi(x_{1})$. Integrating (\ref{*A.15}) with respect to $x_{2}$ we get
\begin{equation}
\partial_{x_{1}}G(a_{1},b_{1},a_{2},b_{2}) = 4 \int_{-\infty}^{x_{2}}\textrm{d}u_{2} \, P(x_{1},u_{2};\rho) + \partial_{x_{1}}G(a_{1},b_{1},a_{2},\sigma_{2}x_{2}) \big\vert_{x_{2}=-\infty} ,
\end{equation}
and performing the integral with respect to $x_{1}$ we find
\begin{eqnarray} \nonumber
G(a_{1},b_{1},a_{2},b_{2}) & = & 4 \Phi(x_{1},x_{2};\rho) + \int_{-\infty}^{x_{1}} \textrm{d}u_{1} \, \partial_{u_{1}}G(a_{1},\sigma_{1}u_{1},a_{2},\sigma_{2}x_{2}) \big\vert_{x_{2}=-\infty} + \\
& + & G(a_{1},\sigma_{1}x_{1},a_{2},b_{2}) \big\vert_{x_{1}=-\infty} ,
\end{eqnarray}
which can be written as
\begin{equation}
G(a_{1},b_{1},a_{2},b_{2}) = 4 \Phi(x_{1},x_{2};\rho) + G(a_{1},-\infty,a_{2},b_{2}) + G(a_{1},b_{1},a_{2},-\infty) - G(a_{1},-\infty,a_{2},-\infty) ;
\end{equation}
the r.h.s. can be further simplified thanks to the integral $I$, thus we get the more transparent expression
\begin{equation}
G(a_{1},b_{1},a_{2},b_{2}) = 4 \Phi(x_{1},x_{2};\rho) - \text{erf}(x_{1}/\sqrt{2}) - \text{erf}(x_{2}/\sqrt{2}) - 1,
\end{equation}
which after a little algebra reduces to
\begin{equation}
G(a_{1},b_{1},a_{2},b_{2}) = 4\Theta(x_{1},x_{2}) + 1 - 4T\left(x_{1},\frac{x_{2}/x_{1}-\rho}{\sqrt{1-\rho^{2}}}\right) - 4T\left(x_{2},\frac{x_{1}/x_{2}-\rho}{\sqrt{1-\rho^{2}}}\right) .
\end{equation}
The latter coincides with (\ref{*A.08}), which is finally proved.


\section{Brownian bridge}
\label{App_B}
A brownian bridge is a Brownian motion constrained to come back to its initial position after a fixed amount of time. We set the initial and final position to be $x=0$, with the motion occurring along the real axis $x$. We consider a set of $n$ infinitesimal space intervals of the form $\mathcal{I}_{j}= (x_{j},x_{j}+\rd x_{j})$ located at times $t_{j}$ with $j\in\{1,\dots,n\}$. The probability for the Brownian path to intersect (pass through) the interval $\mathcal{I}_{j}$ at time $t_{j}$ for \emph{each} $j$ will be $P_{n}(x_{1},t_{1};x_{2},t_{2};\dots;x_{n},t_{n}) \rd x_{1}\rd x_{2}\dots\rd x_{n}$, where $P_{n}(x_{1},t_{1};x_{2},t_{2};\dots;x_{n},t_{n})$ is the joint probability density, which can be deduced on general grounds. Let $W(x_{1},t_{1} \vert x_{0}, t_{0})$ be the transition probability\footnote{See e.g. \cite{Gardiner,VanKampen} for an introduction to stochastic processes.}, which for a Brownian motion takes the well known form
\begin{equation}
\label{*B.01}
W(x_{1},t_{1} \vert x_{0}, t_{0}) = \frac{1}{\sqrt{4\pi D \left(t_{1}-t_{0}\right)}} \text{e}^{-\frac{\left(x_{1}-x_{0}\right)^{2}}{4D\left(t_{1}-t_{0}\right)}} ,
\end{equation}
where $D$ is a constant of diffusion. The probability (\ref{*B.01}) solves the diffusion equation for a Brownian particle which is placed in position $x_{0}$ at time $t_{0}$. Let us consider the case of a single interval for which we can write
\begin{equation}
\label{*B.02}
P_{1}(x,t) = \frac{W(0,T \vert x,t)W(x,t \vert 0,0)}{W(0,T \vert 0,0)} = \sqrt{\frac{T}{4\pi D t \left(T-t\right)}} \text{e}^{-\frac{T}{t(T-t)}\frac{x^{2}}{4D}} .
\end{equation}
Since (\ref{*B.01}) satisfies $\int_{\mathbb{R}}\textrm{d}x \, W(0,T \vert x,t)W(x,t \vert 0,0) = W(0,T \vert 0,0)$, $P_{1}$ is correctly normalized, $\int_{\mathbb{R}}\rd u \, P_{1}(u,t)=1$.
In order to make contact with our notations for phase separation, we write
\begin{equation}
\label{*B.03}
\frac{t_{j}}{T}=\frac{y_{j}}{R}+\frac{1}{2}=\frac{1+\epsilon_{j}}{2};
\end{equation}
it is then simple to see that (\ref{*B.02}) becomes exactly (\ref{*1.12}) provided a suitable identification of the diffusion coefficient is chosen, namely $DT=\lambda^{2}$, with $\lambda$ given by (\ref{etaeps}). 

We consider now the case of $n=2$ intervals. The joint probability distribution for the passage in the intervals $(x_{1},x_{1}+\rd x_{1})$ at time $t_{1}$ and $(x_{2},x_{2}+\rd x_{2})$ at time $t_{2}<t_{1}$ is given by
\begin{equation}
\label{*B.04}
P_{2}(x_{1},t_{1};x_{2},t_{2}) = \frac{W(0,T \vert x_{1},t_{1})W(x_{1},t_{1} \vert x_{2},t_{2})W(x_{2},t_{2} \vert 0,0)}{W(0,T \vert 0,0)} ;
\end{equation}
using (\ref{*B.03}) we find that (\ref{*B.04}) coincides with the joint passage probability (\ref{*4.07}). It is understood that the time ordering is the one depicted in Fig.\ref{fig.04}.

The Brownian properties of interfaces in two dimensions has been investigated with matematically rigorous methods in \cite{GI,CIV}.

\section{Computational toolbox}
\label{App_C}
In this appendix we itemize the functions $\tau_{\Delta,j}$ needed for the computations presented in Sec.\ref{*sec.06}. For each of these functions we provide the corresponding integral over the plane $(\eta_{1},\eta_{2})$ that, according to (\ref{*7.14}), we denote by $\llbracket \tau_{\Delta,j} \rrbracket$. Then we also list the corresponding Fourier-like integrals $\widehat{\llbracket \tau_{\Delta,j} \rrbracket}$ defined by (\ref{*7.15}). The results are 
\begin{align}
\label{*C.01}
\tau_{-2,1}(\theta_{1},\theta_{2},\theta_{3}) &= \frac{1}{\theta_{13}\theta_{32}} , 										&	\llbracket \tau_{-2,1} \rrbracket & = -2\pi^{5/2} \left(1-|\epsilon|\right)^{2} , \\
\label{*C.02}
\tau_{0,1}(\theta_{1},\theta_{2},\theta_{3}) &= 1 ,															&	\llbracket \tau_{0,1} \rrbracket & = 4\pi^{5/2} , \\
\label{*C.03}
\tau_{0,2}(\theta_{1},\theta_{2},\theta_{3}) & = \frac{\theta_{1}^{2}+\theta_{2}^{2}}{\theta_{13}\theta_{32}} , 				&	\llbracket \tau_{0,2} \rrbracket & = 2\pi^{5/2} \left(1+2|\epsilon|-3\epsilon^{2}\right) , \\
\label{*C.04}
\tau_{2,1}(\theta_{1},\theta_{2},\theta_{3}) & = \theta_{1}^{2}+\theta_{2}^{2}, 										&	\llbracket \tau_{2,1} \rrbracket & = 4\pi^{5/2} , \\
\label{*C.05}
\tau_{2,2}(\theta_{1},\theta_{2},\theta_{3}) & = \frac{\theta_{1}^{4}+\theta_{2}^{4}}{\theta_{13}\theta_{32}}, 					&	\llbracket \tau_{2,2} \rrbracket & = 3\pi^{5/2}(3+2|\epsilon|-5\epsilon^{2}) , \\
\label{*C.06}
\tau_{2,3}(\theta_{1},\theta_{2},\theta_{3}) & = \frac{\theta_{1}^{2}\theta_{2}^{2}}{\theta_{13}\theta_{32}}, 					&	\llbracket \tau_{2,3} \rrbracket & = \frac{\pi^{5/2}}{2}(-7+6|\epsilon|-15\epsilon^{2}) ,
\end{align}

\begin{eqnarray}
\label{*C.07}
\pi^{-5/2}\widehat{\llbracket \tau_{-2,1} \rrbracket} & = & -\frac{8}{Q^{2}}\Bigl( 1-\varphi_{1}(Q) \Bigr) , \\
\label{*C.08}
\pi^{-5/2}\widehat{\llbracket \tau_{0,1} \rrbracket} & = & 8\varphi_{1}(Q) , \\
\label{*C.09}
\pi^{-5/2}\widehat{\llbracket \tau_{0,2} \rrbracket} & = & -8\varphi_{2}(Q) , \\
\label{*C.10}
\pi^{-5/2}\widehat{\llbracket \tau_{2,1} \rrbracket} & = & 8\varphi_{1}(Q) , \\
\label{*C.11}
\pi^{-5/2}\widehat{\llbracket \tau_{2,2} \rrbracket} & = & -12\varphi_{3}(Q) , \\
\label{*C.12}
\pi^{-5/2}\widehat{\llbracket \tau_{2,3} \rrbracket} & = & -16\varphi_{1}(Q)-6\varphi_{3}(Q) ,
\end{eqnarray}
where we defined
\begin{eqnarray}
\label{*C.13}
\varphi_{1}(Q) & = & \frac{\sin Q}{Q} , \\
\label{*C.14}
\varphi_{2}(Q) & = & \frac{Q+2Q\cos Q - 3\sin Q}{Q^{3}} , \\
\label{*C.15}
\varphi_{3}(Q) & = & \frac{Q+4Q\cos Q - 5\sin Q}{Q^{3}} .
\end{eqnarray}

The detailed computation of (\ref{*C.01}-\ref{*C.06}) can be quite tedious. We illustrate it through the example of $\llbracket \tau_{-2,1} \rrbracket$, for which we have
\begin{equation}
\label{*C.26}
\llbracket \tau_{-2,1} \rrbracket = \lim_{\Lambda\rightarrow\infty}\int_{-\Lambda}^{\Lambda}\textrm{d}\eta_{1}\int_{-\Lambda}^{\Lambda}\textrm{d}\eta_{2} \bigl\{ \tau_{-2,1} \bigr\} \Bigr\vert_{\epsilon \rightarrow |\epsilon|}.
\end{equation}
Eqs. (\ref{*3.09}), (\ref{*3.10}) and (\ref{*7.13}) imply $\bigl\{ \tau_{-2,1} \bigr\} = - \pi^{5/2} \mathcal{G}(\eta_{1},\epsilon;\eta_{2},-\epsilon,)$, while from (\ref{*3.21}) $\mathcal{G}(\eta_{1},\epsilon;\eta_{2},-\epsilon) = \text{sign}(\eta_{1}\eta_{2}) - 4T(\sqrt{2}\chi_{1},q_{1}) - 4T(\sqrt{2}\chi_{2},q_{2})$ with $\chi_{j} = \frac{\eta_{j}}{\sqrt{1-\epsilon^{2}}}$ and
\begin{eqnarray}
\label{*C.28}
q_{1} & = & \frac{1+\epsilon}{2\sqrt{\epsilon}} \frac{\eta_{2}}{\eta_{1}} - \frac{1-\epsilon}{2\sqrt{\epsilon}} , \\
\label{*C.29}
q_{2} & = & \frac{1+\epsilon}{2\sqrt{\epsilon}} \frac{\eta_{1}}{\eta_{2}} - \frac{1-\epsilon}{2\sqrt{\epsilon}} .
\end{eqnarray}
With the rescaling of the integration variables $\eta_{j}=\kappa\chi_{j}$ we find
\begin{equation}
\label{*C.41}
\llbracket \tau_{-2,1} \rrbracket = - \kappa^{2}\pi^{5/2}\lim_{\Lambda\rightarrow\infty}\int_{-\Lambda}^{\Lambda}\textrm{d}\chi_{1}\int_{-\Lambda}^{\Lambda}\textrm{d}\chi_{2} \, \mathcal{G}(\eta_{1},\epsilon;\eta_{2},-\epsilon) ,
\end{equation}
then we note that
\begin{equation}
\label{*C.42}
\partial_{\epsilon}\mathcal{G}(\eta_{1},\epsilon;\eta_{2},-\epsilon) = -\frac{2}{\pi\sqrt{\epsilon}(1+\epsilon)} \text{e}^{-\frac{1+\epsilon}{4\epsilon}\bigl[(\chi_{1}-\chi_{2})^{2}+\epsilon(\chi_{1}+\chi_{2})^{2}\bigr]} \equiv \Pi(\chi_{1},\chi_{2};\epsilon),
\end{equation}
and that
\begin{equation}
\label{*C.43}
\int_{\mathbb{R}^{2}}\textrm{d}\chi_{1}\textrm{d}\chi_{2} \, \Pi(\chi_{1},\chi_{2};\epsilon) = - \frac{4}{(1+\epsilon)^{2}} ;
\end{equation}
since the integral (\ref{*C.41}) vanishes\footnote{For $\epsilon=1$ we use the fact that $\varrho(x,y) = T(\sqrt{2}x,y/x) + T(\sqrt{2}y,x/y)$ fulfills the symmetries $\varrho(x,y) + \varrho(-x,y) = 0$ and $\varrho(x,-y) + \varrho(x,-y) = 0$.} for $\epsilon=1$ we can write
\begin{equation}
\label{*C.44}
\llbracket \tau_{-2,1} \rrbracket = 4\kappa^{2}\pi^{5/2} \int_{1}^{|\epsilon|} \frac{\textrm{d}\epsilon^{\prime}}{(1+\epsilon^{\prime})^{2}}  = -2\pi^{5/2} (1-|\epsilon|)^{2} ,
\end{equation}
and this proves the identity (\ref{*C.01}). 

\section{Correlation function beyond leading order}
\subsection{Field theoretical derivation}
\label{app_correlator}
Here we obtain the first subleading correction of the two-point function within the large $mR$ expansion. The two-point function can be expanded as stated by (\ref{*7.01}); in the present computation we are not restricting our attention to the connected part, thus we drop the superscript CP and write
\begin{equation}
\label{D1}
\langle\sigma_{1}(x_{1};y)\sigma_{2}(x_{2};-y)\rangle_{ab} = \sum_{\ell=0}^{\infty} \bigl[ \langle\sigma_{1}\sigma_{2}\rangle \bigl]_{\ell} (mR)^{-\ell/2}\,,
\end{equation}
with leading term corresponding to $\ell=0$ given by (\ref{*3.16}). Concerning the first correction ($\ell=1$), let us start by considering the connected part, which ultimately is given by the second term on the r.h.s. of (\ref{*7.13}),
\begin{equation}
\label{ }
\bigl[ \langle\sigma_{1}\sigma_{2}\rangle^{\textrm{CP}} \bigl]_{1} =  \biggl\{\frac{\sqrt{2}Q_{-1}}{4\pi^{5/2}|a_{0}|^{2}}\biggr\};
\end{equation}
$Q_{-1}$ can be readily obtained by expanding (\ref{*7.07}) at small rapidities, and a simple calculation gives
\begin{equation}
\label{*7.21}
Q_{-1} = ic_{0}^{(2)} \Delta\langle\sigma_{1}\rangle\tau_{-1,1} + ic_{0}^{(1)}\Delta\langle\sigma_{2}\rangle\tau_{-1,2} ,
\end{equation}
where the superscript $j$ in $c_{0}^{(j)}$ refers to $\sigma_j$, while $\tau_{-1,j}$ are the functions
\begin{eqnarray} \nonumber
\label{*7.22}
\tau_{-1,1}(\theta_{1},\theta_{2},\theta_{3}) & = & \frac{1}{\theta_{13}} , \\ \nonumber
\tau_{-1,2}(\theta_{1},\theta_{2},\theta_{3}) & = & \frac{1}{\theta_{32}} .
\end{eqnarray}
The corresponding scaling functions can be computed with a straightforward calculation and we find
\begin{eqnarray} \nonumber
\label{*7.24}
\bigl\{ \tau_{-1,1} \bigr\} & = & \frac{2\pi^{2}i}{\kappa} \text{e}^{-\chi_{2}^{2}} \text{erf}\left( \chi_{+} \right) , \\
\bigl\{ \tau_{-1,2} \bigr\} & = & -\frac{2\pi^{2}i}{\kappa} \text{e}^{-\chi_{1}^{2}} \text{erf}\left( \chi_{-} \right) ,
\end{eqnarray}
where
\begin{equation}
\label{*4.19}
\chi_{\pm} = \frac{(1\pm\epsilon)\chi_{1} - (1\mp\epsilon)\chi_{2}}{2\sqrt{\epsilon}} .
\end{equation}
Recalling (\ref{*3.16}), the two-point function decomposes as
\begin{equation}
\label{*7.18}
\langle\sigma_{1}\sigma_{2}\rangle_{ab}\simeq \langle\sigma_{1}\sigma_{2}\rangle_{ab}^{\textrm{CP}} + \widetilde{\langle\sigma_{1}\rangle} \langle\sigma_{2}\rangle_{ab}^{\textrm{CP}} +  \widetilde{\langle\sigma_{2}\rangle} \langle\sigma_{1}\rangle_{ab}^{\textrm{CP}} + \widetilde{\langle\sigma_{1}\rangle}\widetilde{\langle\sigma_{2}\rangle} ,
\end{equation}
where again the superscript CP refers to the contributions coming from the connected part of the matrix element of the order parameter field; in particular we have 
\begin{equation}
\label{*7.19}
\langle\sigma_{j}\rangle_{ab}^{\textrm{CP}} = - \frac{\Delta\langle\sigma_{j}\rangle}{2} \textrm{erf}(\chi_{j}) + c_{0}^{(j)} \frac{P_{1}(x_{j};y_{j})}{m} + \mathcal{O}\left( R^{-1} \right)\,,
\end{equation}
where the second term was determined in \cite{DV} ($P_1$ is given by (\ref{*1.12})). Summing up these findings we obtain the first term beyond leading order in the expansion (\ref{D1})
\begin{eqnarray} \nonumber
\label{*7.26}
\bigl[ \langle\sigma_{1}\sigma_{2}\rangle \bigl]_{1} (mR)^{-1/2} & = & c_{0}^{(1)} \frac{P_{1}(x_{1};y)}{m} \biggl[ \widetilde{\langle\sigma_{2}}\rangle + \frac{\Delta\langle\sigma_{2}\rangle}{2} \text{erf}\left(\chi_{-}\right) \biggr] + \\
& + & c_{0}^{(2)} \frac{P_{1}(x_{2};y)}{m} \biggl[ \widetilde{\langle\sigma_{1}}\rangle - \frac{\Delta\langle\sigma_{1}\rangle}{2} \text{erf}\left(\chi_{+}\right) \biggr] \equiv \mathcal{X}_{1}(x_{1},y;x_{2},-y) .
\end{eqnarray}
It is now rather easy to prove the following clustering relations:
\begin{eqnarray} \nonumber
\label{*7.28}
\lim_{x_{1}\rightarrow + \infty} \mathcal{X}_{1}(x_{1},y;x_{2},-y) & = & c_{0}^{(2)} \frac{P_{1}(x_{2};y)}{m} \langle\sigma_{1}\rangle_{b} , \\ \nonumber
\lim_{x_{1}\rightarrow - \infty} \mathcal{X}_{1}(x_{1},y;x_{2},-y) & = & c_{0}^{(2)} \frac{P_{1}(x_{2};y)}{m} \langle\sigma_{1}\rangle_{a} , \\ \nonumber
\lim_{x_{2}\rightarrow + \infty} \mathcal{X}_{1}(x_{1},y;x_{2},-y) & = & c_{0}^{(1)} \frac{P_{1}(x_{1};y)}{m} \langle\sigma_{2}\rangle_{b} , \\
\lim_{x_{2}\rightarrow - \infty} \mathcal{X}_{1}(x_{1},y;x_{2},-y) & = & c_{0}^{(1)} \frac{P_{1}(x_{1};y)}{m} \langle\sigma_{2}\rangle_{a} .
\end{eqnarray}
which are the counterpart of (\ref{*3.20}) beyond the leading order.

\subsection{Probabilistic interpretation}
\label{app_correlator_prob}
We now show how the correction to the two-point function determined from field theory in the previous section can be interpreted within the framework of section~\ref{*sec.04} endowing the interface with an internal structure. This is done adding to (\ref{Gamma}) the contribution
\begin{equation}
\label{*4.15}
\Gamma_{ab}^{(s)} = \mathscr{A}_{1}^{(0)}\delta(u_{1}-x_{1}) \mathscr{S}_{ab}(x_{2}-u_{2}) + \mathscr{A}_{2}^{(0)}\delta(u_{2}-x_{2}) \mathscr{S}_{ab}(x_{1}-u_{1}) + \dots ,
\end{equation}
where $\mathscr{S}_{ab}$ is the sharp interface profile given by (\ref{*6.03}), and $\mathscr{A}_{1}^{(0)},\mathscr{A}_{2}^{(0)}$ are constants which, due to the delta functions, carry information about a structure located {\it on} the interface. The correction to (\ref{*4.01}) coming from this modification of (\ref{Gamma}) is
\begin{equation}
\langle\sigma_{1}(x_{1},y)\sigma_{2}(x_{2},-y)\rangle_{ab}^{(1)} = \int_{\mathbb{R}^{2}}\textrm{d}u_{1}\textrm{d}u_{2} \, P_{2}(u_{1},y;u_{2},-y) \Gamma_{ab}^{(s)}(x_{1},y;x_{2},-y\vert u_{1},u_{2}) ,
\end{equation}
which after simple manipulations it becomes
\begin{eqnarray} \nonumber
\label{*4.16}
\langle\sigma_{1}(x_{1},y)\sigma_{2}(x_{2},-y)\rangle_{ab}^{(1)} & = & \mathscr{A}_{1}^{(0)} \langle\sigma_{2}\rangle_{a} \int_{x_{2}}^{+\infty}\textrm{d}u_{2} \, P_{2} + \mathscr{A}_{1}^{(0)} \langle\sigma_{2}\rangle_{b} \int_{-\infty}^{x_{2}}\textrm{d}u_{2} \, P_{2} + \\ \nonumber
& + & \mathscr{A}_{2}^{(0)} \langle\sigma_{1}\rangle_{a} \int_{x_{1}}^{+\infty}\textrm{d}u_{1} \, P_{2} + \mathscr{A}_{2}^{(0)} \langle\sigma_{1}\rangle_{b} \int_{-\infty}^{x_{1}}\textrm{d}u_{1} \, P_{2} ,
\end{eqnarray}
where $P_{2}$ stands for $P_{2}(x_{1},y;x_{2};-y)$. After a rescaling of the integration variables we can cast the above in the form
\begin{eqnarray} \nonumber
\label{*4.17}
\langle\sigma_{1}(x_{1},y)\sigma_{2}(x_{2},-y)\rangle_{ab}^{(1)} & = & \mathscr{A}_{1}^{(0)} \lambda^{-1} \langle\sigma_{2}\rangle_{a} \mathcal{W}_{2}^{+} + \mathscr{A}_{1}^{(0)} \lambda^{-1} \langle\sigma_{2}\rangle_{b} \mathcal{W}_{2}^{-} + \\
& + & \mathscr{A}_{2}^{(0)} \lambda^{-1} \langle\sigma_{1}\rangle_{a} \mathcal{W}_{1}^{+} + \mathscr{A}_{2}^{(0)} \lambda^{-1} \langle\sigma_{1}\rangle_{b} \mathcal{W}_{1}^{-} ,
\end{eqnarray}
where $\mathcal{W}_{j}^{\pm}=\mathcal{W}_{j}^{\pm}(\eta_{1},\eta_{2};\epsilon)$ are the functions
\begin{eqnarray} \nonumber
\label{*4.18}
\mathcal{W}_{1}^{+} & = & \int_{\eta_{1}}^{+\infty}\textrm{d}h_{1} \, U(h_{1},\eta_{2};\epsilon) = \frac{\textrm{e}^{-\chi_{2}^{2}}}{2\sqrt{\pi}\kappa} \biggl[ 1 - \textrm{erf}(\chi_{+}) \biggr] , \\ \nonumber
\mathcal{W}_{1}^{-} & = & \int_{-\infty}^{\eta_{1}}\textrm{d}h_{1} \, U(h_{1},\eta_{2};\epsilon) = \frac{\textrm{e}^{-\chi_{2}^{2}}}{2\sqrt{\pi}\kappa} \biggl[ 1 + \textrm{erf}(\chi_{+}) \biggr] , \\ \nonumber
\mathcal{W}_{2}^{+} & = & \int_{\eta_{2}}^{+\infty}\textrm{d}h_{2} \, U(\eta_{1},h_{2};\epsilon) = \frac{\textrm{e}^{-\chi_{1}^{2}}}{2\sqrt{\pi}\kappa} \biggl[ 1 + \textrm{erf}(\chi_{-}) \biggr] , \\ \nonumber
\mathcal{W}_{2}^{-} & = & \int_{-\infty}^{\eta_{2}}\textrm{d}h_{2} \, U(\eta_{1},h_{2};\epsilon) = \frac{\textrm{e}^{-\chi_{1}^{2}}}{2\sqrt{\pi}\kappa} \biggl[ 1 - \textrm{erf}(\chi_{-}) \biggr] , \\
U(\eta_{1},\eta_{1};\epsilon) & = & \textrm{e}^{- \frac{\eta_{1}^{2}+\eta_{2}^{2}}{2(1-\epsilon)} - \frac{(\eta_{1}-\eta_{2})^{2}}{4\epsilon}} .
\end{eqnarray}
Therefore using the known expression (\ref{*1.12}) for the passage probability $P_{1}$ and the functions $\mathcal{W}_{j}^{\pm}$, (\ref{*4.17}) finally becomes
\begin{eqnarray} \nonumber
\label{*4.20}
\langle\sigma_{1}(x_{1},y)\sigma_{2}(x_{2},-y)\rangle_{ab}^{(1)} & = & \mathscr{A}_{1}^{(0)} P_{1}(x_{1},y) \biggl[ \widetilde{\langle\sigma_{2}\rangle} + \frac{\Delta\langle\sigma_{2}\rangle}{2}\textrm{erf}(\chi_{-}) \biggr] + \\
& + & \mathscr{A}_{2}^{(0)} P_{1}(x_{2},-y) \biggl[ \widetilde{\langle\sigma_{1}\rangle} - \frac{\Delta\langle\sigma_{1}\rangle}{2}\textrm{erf}(\chi_{+}) \biggr] .
\end{eqnarray}
This coincides with the field theoretical result (\ref{*7.26}) once one identifies $\mathscr{A}_{j}^{(0)}=c_{0}^{(j)}/m$.

Lastly, we comment on the terms omitted in (\ref{*4.15}). We notice that (\ref{Gamma}) can be written in the compact form 
\begin{equation}
\label{*4.21}
\mathscr{S}_{ab}(x_{1}-u_{1})\mathscr{S}_{ab}(x_{2}-u_{2}) ,
\end{equation}
and that this suggest the factorized expression
\begin{equation}
\label{*4.22}
\Gamma_{ab}(x_{1},y_{1};x_{2},y_{2} \vert u_{1},u_{2}) = \sigma_{ab}(x_{1} \vert u_{1}) \sigma_{ab}(x_{2} \vert u_{2})
\end{equation}
for the function entering (\ref{*4.01}); here 
\begin{equation}
\label{*4.23}
\sigma_{ab}(x_{j} \vert u_{j}) = \mathscr{S}_{ab}(x_{j}-u_{j}) + \mathscr{A}_{j}^{(0)}\delta(x_{j}-u_{j}) + \mathscr{A}_{j}^{(1)}\delta^{\prime}(x_{j}-u_{j}) + \mathscr{A}_{j}^{(2)}\delta^{\prime\prime}(x_{j}-u_{j}) + \dots ,
\end{equation}
where the prime symbol stands for the derivative with respect to $u_{j}$. Eq. (\ref{*4.23}) is exactly the sharp profile dressed with local terms accounting for interfacial structure proposed in \cite{DV} within the study of the one-point function.


\end{document}